\documentclass[useAMS,usenatbib]{mnras}
\usepackage{amsmath}
\usepackage{graphicx,txfonts,fleqn,enumerate,color}

\renewcommand*{\v}[1]  {\boldsymbol{#1}}

\newcommand*  {\twovector}[2] {{\begin{pmatrix} $1 \\ $2 \end{pmatrix}}}

\renewcommand {\emph}[1]  {\textit{#1}}

\title[Are long-term $N$-body simulations reliable?]
{{Are long-term $N$-body simulations reliable?}}

\author[David M. Hernandez, Sam Hadden, and Junichiro Makino]
	{David M. Hernandez$^{1,2}$\thanks{{{Email: dmhernandez@cfa.harvard.edu} }}, Sam Hadden$^{1}$, and Junichiro Makino$^{2}$  \\ 
	$^1$ Harvard--Smithsonian Center for Astrophysics, 60 Garden St., MS 51, Cambridge, MA 02138, USA \\
	$^2$ {RIKEN Center for Computational Science}, 7-1-26 Minatojima-minami-machi, Chuo-ku, Kobe, 650-0047 Hyogo, Japan \\
	}
\begin{document}

\maketitle

\label{first page}
\begin{abstract} 
$N$-body integrations are used to model a wide range of astrophysical dynamics, but they suffer from errors which make their orbits diverge exponentially in time from the correct orbits.  Over long time-scales, their reliability needs to be established.  We address this reliability by running a three-body planetary system over about $200$ e-folding times.  Using nearby initial conditions, we can construct statistics of the long-term phase-space structure and compare to rough estimates of resonant widths of the system.  {We compared statistics for a} wide range of numerical methods, including a Runge--Kutta method, Wisdom--Holman method, symplectic corrector methods, and a method by Laskar and Robutel. ``Improving'' an integrator did not {increase} the phase space accuracy, but simply increasing the number of initial conditions did.  {In fact, the statistics of a higher order symplectic corrector method were inconsistent with the other methods in one test.}
\end{abstract}
\section{Introduction}
\label{sec:first}

In the gravitational $N$-body problem \citep{HH03}, $N$ particles interact with each other through an inverse square force law.  Corrections can be applied to this model \citep{Tamayoetal2019}, but this simple model alone is used to represent dynamical systems from planetary to cosmological.  Due to the inverse square law, far away particles in self-gravitating systems are coupled, unlike the situation for more steep potentials.  Due to the Coulomb logarithm \citep{BT08}, particles between $10^n$ and $10^{n+1}$ units of length in separation contribute the same to changing the mean-square velocity as particles with separations between $10^{n+1}$ and $10^{n+2}$, assuming a system of fixed density.

The $N$-body equations are solved using approximate mappings called integrators.  Alarmingly little is known about the accuracy of these maps.  They suffer from truncation and roundoff error, and error due to the uncertainty in the initial conditions.  \cite{Miller1964} found that the orbits between a system and one that is slightly perturbed to it diverge exponentially so the errors can magnify impressively in short times.  {The exponential divergence is a property of the $N$-body equations.}  Thus, the orbits of one system have no resemblance to the orbits in the other system.  However, $N$-body integrations are a standard tool for astrophysicists, despite there being a limited justification for their use.

Fast symplectic \citep{hair06,Y90} integrators that make use of the near integrability of the system, are standard tools for the exploration of planetary dynamics \citep{WH91,Kinoshitaetal91}.  These integrators have been improved significantly and continue to be in development.  But are smaller errors better, or are the final predictions just as unreliable?  Are symplectic maps more reliable than conventional integrators, which have been tested, and have been in existence for far longer?  No answer is known to these questions in general.

However, for investigations over short timescales, an answer is known to these questions.  By short or long-term we refer to a small or large number number of e-folding times.  In this category, we also include regular problems with no exponential divergence, even if they are simulated for a large number of orbits, because the same arguments apply to them.  These definitions are arguably unconventional, but we find them more useful for studying integration errors.  \cite{Reinetal2019c} study the Solar System for about $4$ e-folding times, a single planet with general relativistic precession, with no exponential divergence, and a test consisting of Mercury, Saturn, and the Sun, which shows no clear exponential divergence in our own tests over $100,000$ yrs.  According to our definitions, their studies are short-term, even though they refer to them as long-term.  An e-folding time is termed a Lyapunov time if the exponential divergence of infinitesimally close phase space points continues indefinitely.  In a globular cluster, for example, the concept of a Lyapunov time is not useful because the asymptotic system has ejected stars and is integrable.  An e-folding time could also be termed a local Lyapunov exponent \citep{QT92}.  

In the short-term regime, conventional integrators can calculate exact orbits with some known error in the orbit.  We assume relative roundoff and truncation errors can be kept small \citep{RS15}.  Fig. 3 of \cite{Wisdom2018} provides evidence that \texttt{IAS15}, a conventional integrator, is just as efficient as any alternative for these purposes in their short-term simulation.  {\cite{Reinetal2019c} study the accuracy of symplectic integrators, but state in their Appendix A that a conventional integrator is better at calculating exact angle variables\footnote{{{The roundoff error of symplectic integrators can also be kept small \citep{Farresetal2013}.  {However, a case in which such integrators can achieve a smaller error} than the conventional integrator IAS15, when working in double precision, has not been shown in publications to our knowledge.}}}

The situation is drastically different for long-term dynamical investigations, which includes the case of the evolution of the Solar System over its lifetime or of galaxies over cosmic times.  The exponential divergence renders calculating exact orbits and their angle variables, such as periastron, impossible over long time-scales due to a computer's finite memory.  Variable precision arithmetic methods \citep{PB14,BP15,PB18} can mitigate this problem for a limited time.  In the long-term regime, the dynamics become irreversible in a practical sense, even when they are formally reversible.  Even codes especially designed to ensure reversibility \citep{PB18,ReinTamayo2018} will fail irreversibility tests eventually.  Note that irreversibility in of itself is not problematic \citep{Heggie91}, although it can be an indication of erroneous dynamics as \cite{H16} found in investigations of \texttt{MERCURY}.  This long-term regime is what we want to explore in this paper, as it governs the secular behavior of many dynamical systems in astrophysics.  We propose investigating this regime in a statistical manner, discarding all information on individual orbits, as some investigators of self-gravitating systems have done previously. A literature review of attempts to asses reliability of $N$-body is given in Appendix \ref{sec:lit}. For alternate possible ways to address this question, see \cite{QT92,Urminsky2010}.

These works do not look at phase space statistics we are interested in, in planetary dynamics, and they are not run for the long timescales we might be interested in.  They also do not address whether symplectic integrators might better calculate converged statistics.  \cite{chan90} argue that symplectic methods are better than conventional integrators to study near-integrable Hamiltonians.  As for general chaotic Hamiltonians, they state (SIAs are symplectic integrator algorithms), ``Though naively one might think that SIAs and standard integrators should be equally good in the chaotic regions, SIAs are also superior there because they avoid the boundaries and the embedded islands of secondary tori." 

This work is concerned with assessing the long-term reliability of $N$-body systems once the memory of the initial conditions has vanished, by studying the accuracy of the phase space structure.  In addition, we want to study whether symplectic integrators better reproduce the phase space structure, and what benefits we might find from using popular higher order symplectic integrators in explorations of long-term planetary dynamics.  As a test problem, we consider the restricted three-body problem.  We consider chaotic, but Hill stable initial conditions, which means close encounters with the perturbing planet are forbidden by zero-velocity curves \citep[e.g.,][]{MurrayDermott99}.  {These } let us use the popular methods of \cite{WH91,WHT96,W06,LR01}.  The particle avoids regular islands of phase space surrounding the elliptic fixed points of some low-order mean motion resonances (MMRs); the most significant islands occur at first-order MMRs. The widths of these regular regions surrounding first-order MMRs roughly agree with the analytic resonance width estimates in \cite{HL18} despite the significant degree of chaos observed over the range of phase-space explored by the particles.  We find that in long-term studies, conventional integrators perform as well as symplectic integrators as far as accuracy is concerned.  The only advantage we find for symplectic integrators is their speed, if they are of second order.  Higher order symplectic integrators showed no advantages over lower order methods.  We agree in spirit with \cite{Smith77}, who states, ``the most efficient use of a limited amount of computer time is to integrate a large number of different examples at relatively low accuracy.'' 

The paper is organized as follows.  Section \ref{sec:methods} gives a brief description of the integrators we use in our study.  We test their implementation on a test problem with long Lyapunov time.  In Section \ref{sec:choosing}, we present and study the restricted three-body problem we consider for the remainder of the paper.  We describe how we will collect statistics using many nearby initial conditions to the fiducial ones to map the phase space structure of this problem.  Section \ref{sec:comp} carries out the statistical experiments and makes comparisons among the different methods.  Different timesteps are considered.  We conclude in Section \ref{sec:conc}.
\section{Methods}
\label{sec:methods}

This section introduces the integrators we use in the experiments.  The $N$-body Hamiltonians we solve have form,
\begin{equation}
\label{eq:hamilt}
H = A + \epsilon B, 
\end{equation}
where $A$ and $B$ are integrable functions and $\epsilon \ll 1$.  $\epsilon \simeq 0.001$ for the Solar System.  Other conventions for writing the Hamiltonian are possible: \cite{HD17} wrote $H = A + B$, where $A \propto \epsilon$ and $B \propto \epsilon^2$.  $A$ is a sum of Kepler problems, while $B$ is the remaining perturbation.  It's possible to write $H$ in different coordinate systems: popular choices are Jacobi, inertial Cartesian, Democratic Heliocentric, and Canonical Heliocentric coordinates \citep{HD17}, each with its own merits.  $H$ can also be split into functions $A$ and $\tilde{B}$, where $\tilde{B}$ is not integrable, but, in order to use our popular integrators, we do not consider them.

We choose Democratic Heliocentric coordinates.  The integrators we use are the Wisdom--Holman method (WH) \citep{Kinoshitaetal91,WH91}, the Wisdom--Holman method with third-order correctors (WHc) \citep{Butcher69,WHT96,W06}, LR, the $SABA_2$ integrator from \cite{LR01}, and a Runge--Kutta--Fehlberg 4-5 method (RK) \citep[Section 16.2]{press02}.

To describe WH, we follow somewhat the notation of \cite{Tamayoetal2019}, though there are a wide array of notations we can use as options.  Let $\mathcal A(a)$ and $\mathcal B(b)$ be phase space maps derived from Hamilton's equations for $A$ and $\epsilon B$, respectively.  $\mathcal A(a)$ and $\mathcal B(b)$ are canonical transformations described by a constants $a$ and $b$.  Let $h$ be a timestep, whose magnitude is usually set to resolve the shortest timescales in the Hamiltonian initial value problem.  The WH canonical transformation is written,
\begin{equation}
\label{eq:WH}
\phi_{\mathrm{WH}}(h) = \mathcal{A}(h/2) \circ \mathcal{B}(h) \circ \mathcal A(h/2),
\end{equation}
where $\circ$ is a composition of operators \citep{Tamayoetal2019}.  During integration, we use $\mathcal{A}(a) \circ \mathcal{A}(a) = \mathcal{A}(2 a)$, which follows since the Poisson bracket of $A$ with itself is $0$.  Physically, $\phi_{\mathrm{WH}}(h)$ approximates the phase space at time $t + h$, where $t$ was the starting time.  Iterating \eqref{eq:WH} gives the orbits.  Note $\mathcal{B}(h)$ is sandwiched between $\mathcal{A}(h/2)$--- its also possible to sandwich $\mathcal{A}(h)$ between $\mathcal{B}(h/2)$, but this gives an energy error about twice as large \citep{HD17}.  The energy error from iterating \eqref{eq:WH} is $\mathcal{O}(\epsilon h^2 + \epsilon^2 h^2)$ \citep{Reinetal2019b}.

WHc applies extra maps $\mathcal {C}(h)$ that improve the order of WH.   $\mathcal {C}(h)$ is a composition of maps, $\mathcal{A}(a_i)$ and $\mathcal{B}(b_i)$, where the $a_i$ and $b_i$ are constants and $i = 1$ or $2$.  The WHc map is 
\begin{equation}
\label{eq:WHc}
\phi_{\mathrm{WHc}}(h) = \mathcal{C}^{-1}(h) \circ \mathcal{A}(h/2) \circ \mathcal{B}(h) \circ \mathcal A(h/2) \circ \mathcal {C}(h),
\end{equation}
where $\mathcal C^{-1}(h) \circ \mathcal C(h) = 1$.  Thus, if we iterate \eqref{eq:WHc}, we only need to apply two correctors, $\mathcal {C}(h)$, which is applied to the initial conditions, and $ \mathcal{C}^{-1}(h)$, applied as the last map (or operator).  The energy error of WHc is $\mathcal O(\epsilon h^4 + \epsilon^2 h^2)$.  Applying correctors in this form has uses such as converting a reversible method to a symplectic one \citep{hair06,HB18}.

LR has the same order energy error as WHc.  It is composed of $\mathcal{A}(a_i)$ and $\mathcal{B}(b_i)$ operators.  The map is,
\begin{equation}
\label{eq:LR}
\phi_{LR}(h) = \mathcal{A}(c_1 h) \circ \mathcal{B}(d_1 h) \circ \mathcal A(c_2 h) \circ \mathcal{B}(d_1 h) \circ \mathcal A(c_1 h),
\end{equation}
where $d_1 = 1/2$, $c_1 = \frac{1}{2} (1- 1/\sqrt{3})$, and $c_2 = 1 - 2 c_1$.  The coefficients $d_1$, $c_1$, and $c_2$ are positive, which has advantages \citep{DH17}.  The order of LR and WHc can be improved by applying extra operators called kicks, such techniques are beyond our concern in this work and are described in many places \citep{Chin97,ChinandChen05,W06,Reinetal2019b,DH17}.  WH, WHc, and LR are symplectic \citep{hair06}, time-reversible \citep{hair06,HB18}, conserve angular and linear momentum, and their corresponding ordinary differential equations (ODEs) have infinite differentiability class order \citep{H19,H19b}, just as the $N$-body Hamiltonian itself.  These features improve error behavior.  A disadvantage is that they assume $\epsilon B \ll A$.  \cite{DH17} mitigates this by removing two-body error terms, and its method has error $\mathcal O(\epsilon h^4)$.

RK is neither symplectic nor time-reversible, but, curiously, has infinite differentiability class order.  It takes adaptive stepsizes based on estimates of the error.  It will serve as a benchmark method against which to compare the other methods.

We test our implementation of WH, WHc, and LR using a planar restricted three-body problem consisting of the ``Sun,'' ``planet,'' and test particle.  The orbit is chaotic but Hill stable \citep{HL18}.  Enforcing Hill stability guarantees $\epsilon B \ll A$.

Letting $G = 1$, the masses are $m_1 = 1$, $m_2 = 3\times 10^{-5}$, and $m_3 = 0$.  The initial Cartesian positions are in an inertial frame centered and at rest. {All particles are aligned {.  In the rotating frame, the test particle initially moves clockwise}.  The test particle is at apocenter with semi-major axis $a \approx 0.21$ and eccentricity $e \approx 0.19$.  The planet has semi-major axis $a^\prime \approx 0.29$.  The exact Cartesian coordinates are given in eqs. \eqref{eq:icscale}. }

The planet's period is $1$\footnote{The integrator modifies this period by $\approx m_2/(2 m_1)$.}.  We run these initial conditions using the three maps and various $h$ for time $t = 100$.  We record $| \Delta J/J |$ at $t = 100$.  The {local} Lyapunov exponent for this problem is longer than the runtime.  We estimate Lyapunov exponents in this paper by running two integrations.  The second one displaces the test-particle's $x$-position by $+10^{-14}$.  We calculate an $L_2$ norm as a function of time:
\begin{equation}
L_2(t) =  \sqrt { |\v{Q}_{3p}(t) - \v{Q}_3(t)|^2 + |\v{v}_{3p}(t) - \v{v}_3(t)|^2 }.
\label{eq:l2}
\end{equation}
$\v{Q}_{3}$ is the unperturbed test particle's heliocentric positions while $\v{Q}_{3p}$ are the positions in the perturbed run.  $\v{v}_{3}$ and $\v{v}_{3p}$ are the velocity vectors.  All tests in this paper compare initial conditions whose test-particle $x$-position is displaced in this way, so using \eqref{eq:l2} to quantify Lyapunov exponents proves useful to us.  Measuring Lyapunov exponents in this way is reliable, except for a set of initial conditions of measure $0$ in the continuous problem. {We verified the long local Lyapunov exponent of this problem using the Mean Exponential Growth factor of Nearby Orbits \citep{Cincottaetal2003} in IAS15 \citep{RS15}. }

The results of our experiments are shown in Fig. \ref{fig:scaling}.  For small steps, the three methods scale as $h^2$.  At larger steps, WHc and LR scale as $h^4$.  For $h \gtrsim 0.05$, the curves get distorted, as $h$ no longer resolves well the timescales of the test-particle orbit.

\begin{figure}
	\includegraphics[width=90mm]{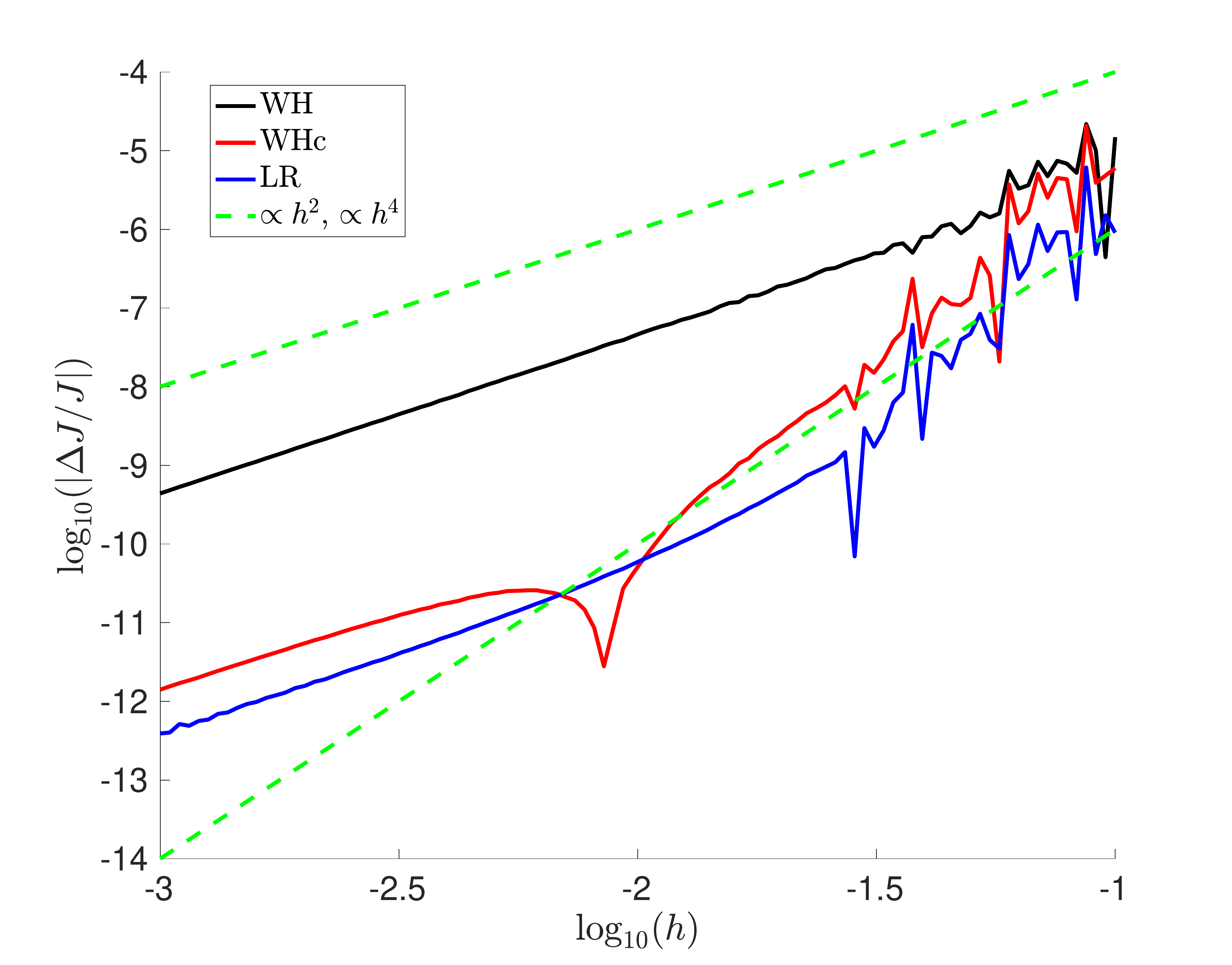}
	\caption{We test our implementation of various symplectic integrators.  We run a planar, restricted three-body problem and record the Jacobi constant error $|\Delta J/J|$ at time $t = 100$ using various timesteps $h$.  For small timesteps, the scaling is $h^2$, but at larger steps, WHc and LR scale as $h^4$.
	\label{fig:scaling}
  	}
\end{figure}

\section{Test problem}
\label{sec:choosing}

We wish to study orbital dynamics over long timescales (cf. Section \ref{sec:first}), and the three-body problem of Section \ref{sec:methods} has too long a Lyapunov time for this purpose.  Its long Lyapunov time allowed a clear scaling study.  We find initial conditions that have shorter Lyapunov time while still remaining Hill stable, and orbit the primary. {They are identical to those of Section \ref{sec:methods}, except that now $a \approx 0.27$ and $e \approx 0.036$.  Note the larger apocenter.  Exact initial conditions are given in eqs. \eqref{eq:ic}.}

{Local} Lyapunov exponents are computed in Table \ref{tab:lyap} using WH with different $h$ {, with a runtime of $t = 150$.}  Note the Lyapunov time varies slightly with $h$.  {Local Lyapunov exponents as a function of time were computed again using IAS15.  We found that between $t = 0$ and $t = 3500$, the local Lyapunov exponent varied between $14$ and $24$, approximately consistent with Table \ref{tab:lyap}.}
\begin{center}
\begin{table*}
\caption{{Local} Lyapunov exponents $t_{\mathrm{Ly}}$ for initial conditions \eqref{eq:ic}.  The exponent varies with timestep $h$ of the WH method used to compute it.}
\centering
\begin{tabular}{| c || c|}
	\hline
	 $h$ & $t_{\mathrm{Ly}}$ \\ [3ex] \hline
	 $0.001$ & $10$ \\ \hline
	$0.01$ & $15$ \\ \hline
	$0.05$ & $19$ \\ \hline
	$0.1$ & $18$ \\ \hline
\end{tabular}
\label{tab:lyap}
\end{table*}
\end{center}
We look at these initial conditions more closely in Fig. \ref{fig:hadden2}.  First, we present some definitions: let $\lambda$ and $\lambda^\prime$ be the test particle and planet mean longitudes, respectively.  $\varpi$ is the longitude of periapsis.
\begin{figure}
	\includegraphics[width=90mm]{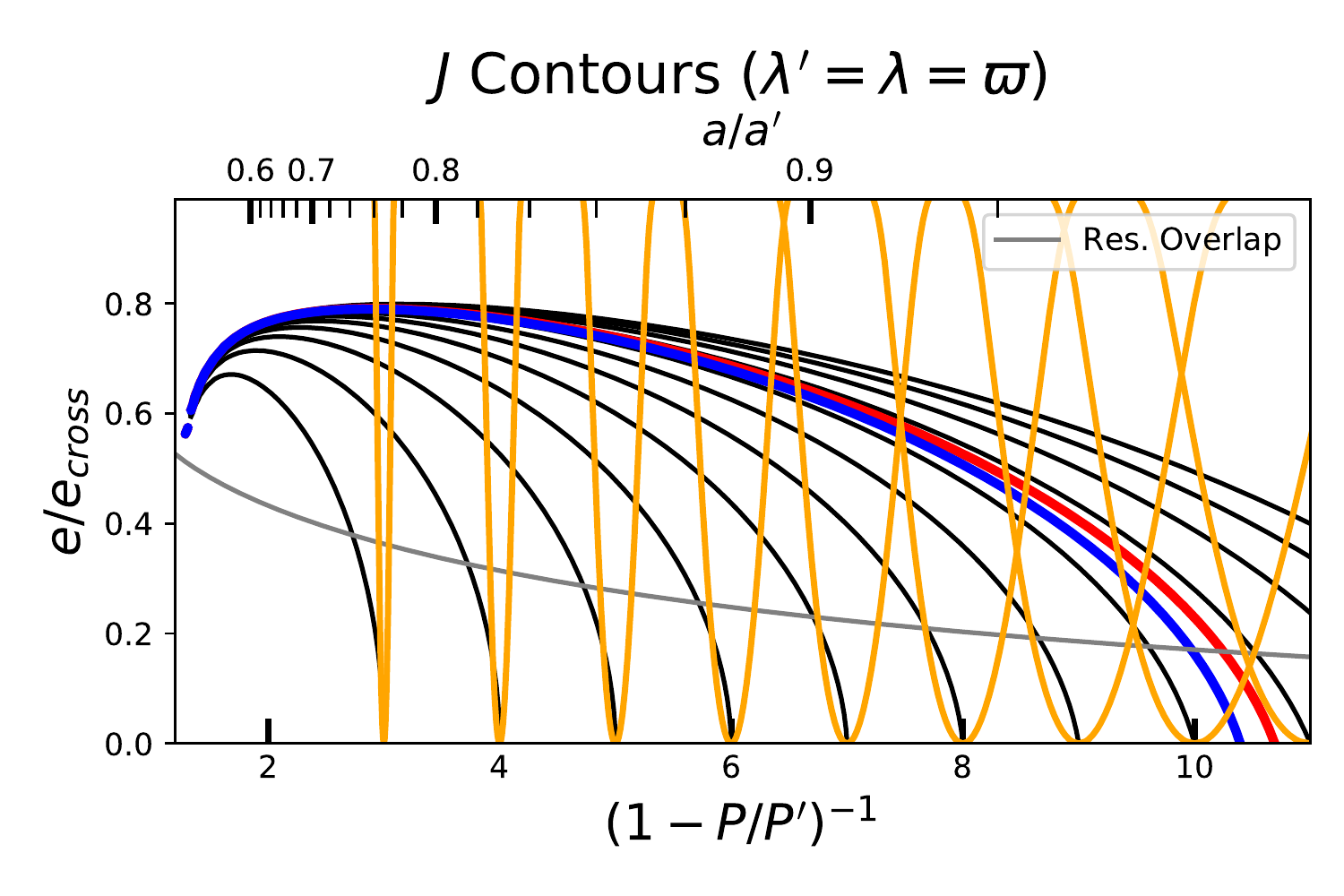}
	\caption{{Jacobi constant contours projected in the plane $(1-P/P')^{-1}$ versus $e/e_\text{cross}$ when the test-particle and perturber are in the configuration $\lambda=\lambda^\prime=\varpi$.  The Jacobi constant of the initial conditions in Equation \eqref{eq:ic} is shown in blue. The Jacobi constant of the $L_1$ Lagrange point, marking the boundary of Hill stability is shown in red. Separatrices of first-order MMRs, estimated according to Equations 5 and 7 of \citet{HL18} are plotted in orange. (Note that the separatrices of first-order resonances have their maximal width as measured in $(1-P/P')^{-1}$ for $\lambda'=\lambda=\varpi$).  The resonance overlap criterion of \citet{HL18} is shown as a gray line.}
	\label{fig:hadden2}
  	}
\end{figure}
Consider the Jacobi constant calculated from eq. \eqref{eq:ic}.  The $x$-axis measures $j = (1 - P/P^\prime)^{-1}$, where $P$ is the initial osculating period of the test-particle, and $P^\prime$ is the planet's period ($= 1$).  $j = i$, with $i$ an integer, indicates an $i:i-1$ mean motion resonance.  The $y$-axis measures the initial osculating eccentricity in units of the crossing eccentricity ($e_{\mathrm{cross}} \equiv \frac{a - a^\prime}{a}$--- note it's not a constant).  $a$ and $a^\prime$ are the osculating semi-major axes.  In yellow are resonance widths, calculated from eq. (5) and (7) from \cite{HL18}.  A gray curve indicates an $e$, as a function of $j$, above which the covering fraction between adjacent $j$ is 1, a criteria for chaos.  The blue curve gives $e(a)$ for the Jacobi constant calculated at conjunction and pericenter.  The red curve indicates the maximum $e(a)$ for Hill-stability.  The initial conditions \eqref{eq:ic} have $a/a^\prime \approx 0.93$ and $j \approx 9.7$.  At this $j$, the Jacobi constant is chaotic but Hill stable and only orbits the primary.  To see this, note that the initial planet-test particle separation is $\approx 0.0106$ while the planet--$L_1$ separation, where $L_1$ is the inner collinear Lagrange point, is $\approx 0.00632$.  Also, the ratio of the $0$-velocity Jacobi constant \citep[Eq. (3.39)]{MurrayDermott99} to the $0$-velocity Jacobi constant at $L_1$ for the Sun-planet configuration is $\approx 0.376$, which is less than $1$, as required: the test-particle is forbidden from crossing $L_1$.

We show the test particle's orbit corresponding to the initial conditions (ICs) of eq. \eqref{eq:ic} in Fig. \ref{fig:traj}.
\begin{figure}
	\includegraphics[width=90mm]{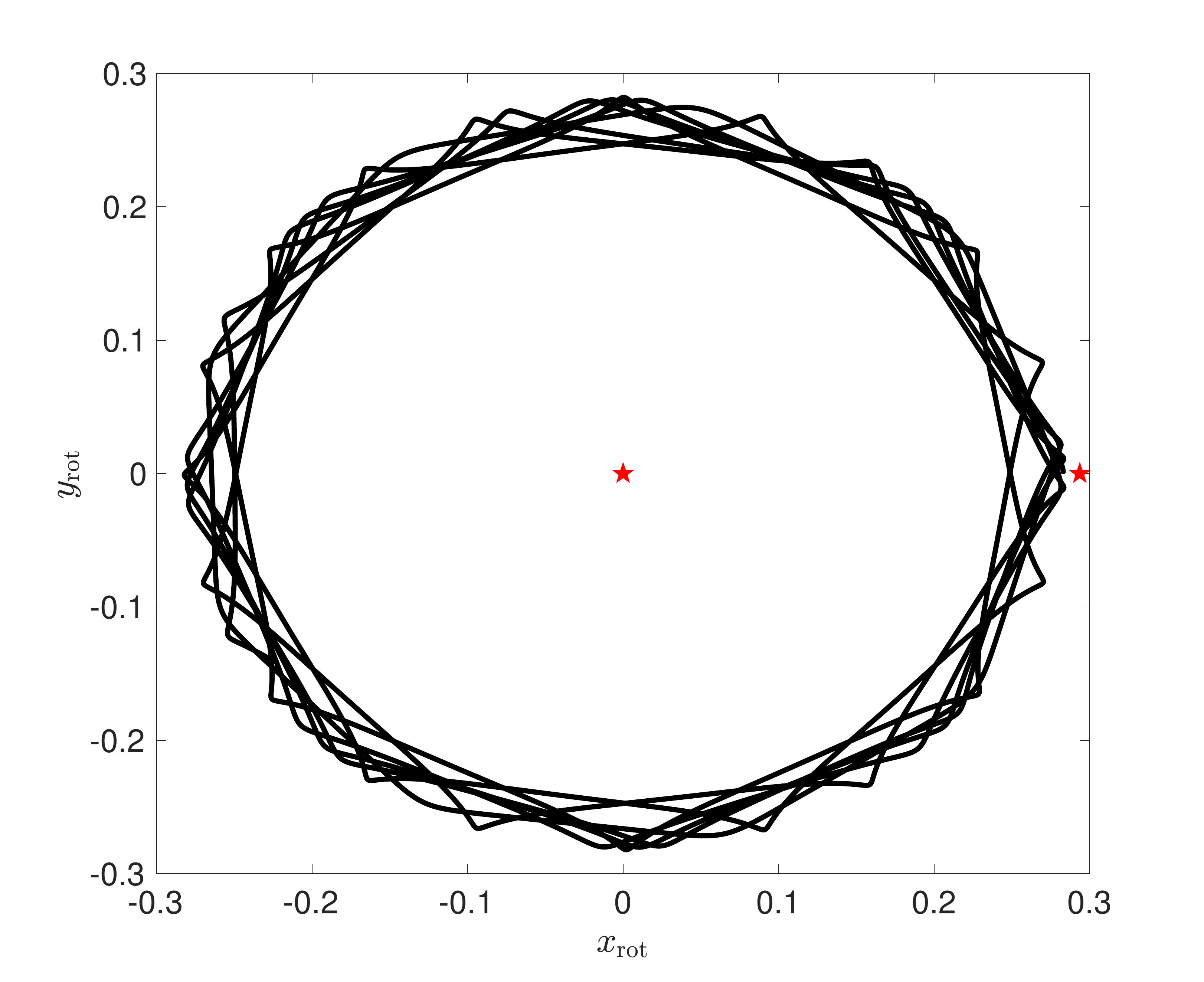}
	\caption{The orbit of the test particle of initial conditions \eqref{eq:ic} in the rotating frame.  The position is plotted up to $t = 50$ and the approximate positions of the Sun and planet are indicated with red stars.
	\label{fig:traj}
  	}
\end{figure}
Its positions in the rotating frame, $x_{\mathrm{rot}}$, $y_{\mathrm{rot}}$, are plotted up to $t = 50$ which is about $3$ Lyapunov times.  The orbit is computed using WH with $h = 0.01$.  The Jacobi error at $t = 50$ is $|\Delta J/J| \approx 7.6 \times 10^{-8}$.  The positions of Sun and planet are shown with red stars.  Note they are approximate since the period is slightly wrong (Section \ref{sec:methods} ).

In Fig. \ref{fig:aevstlong}, $a$ and $e$, calculated as osculating orbital elements are plotted as function of time using WH with $h = 0.01$.  This is a long-term integration, spanning about $200$ Lyapunov times.  Memory of the initial conditions can be said to be lost if, in double precision calculations, an error of $10^{-16}$ in the $L_2$ norm (eq. \eqref{eq:l2}) grows to approximately order unity.  This occurs at roughly $36$ Lyapunov times.  IC 0 refers to initial conditions \eqref{eq:ic} while in IC 1, the initial conditions have been perturbed: the $x$-coordinate of the test-particle has been displaced by $+10^{-14}$.  The evolution of $a$ and $e$ is clearly distinct.  $t = 1500$ is indicated by a thick blue line.  It indicates $75$--$150$ Lyapunov times, larger than the $36$ required for memory loss of the initial conditions.  We will use $t = 1500$, for safety, as the time when memory loss occurs.  After this time, we have confidence that memory of the initial conditions has been lost.  Some values of $a$ and $e$ are preferred, we explain this below.
\begin{figure}
	\includegraphics[width=90mm]{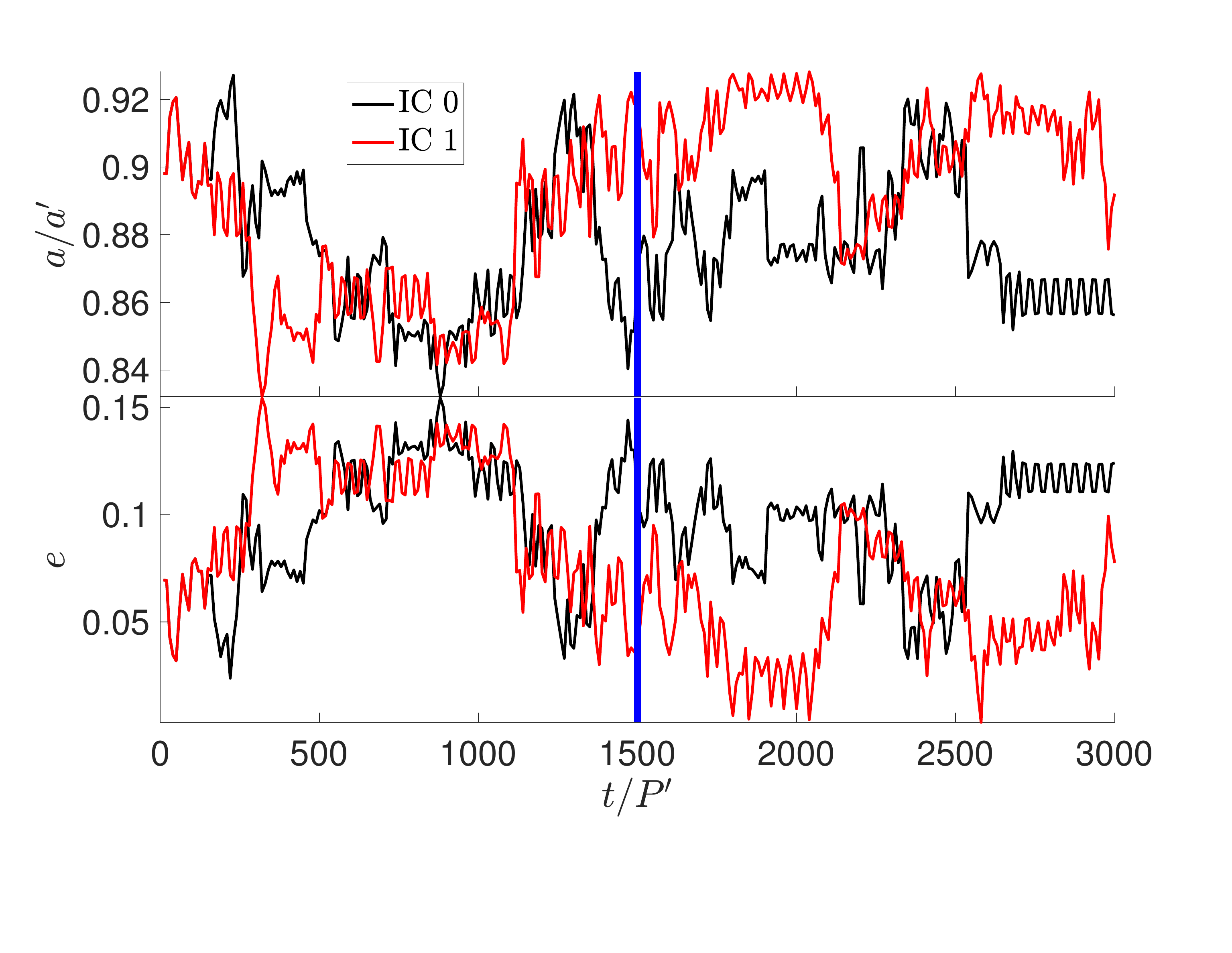}
	\caption{The osculating orbital elements $a$ and $e$, of two test particles with nearby initial conditions.  Some orbital element values are preferred.  A blue line indicates $t = 1500$: we designate it as a time for which memory loss of the initial conditions has occurred.  
	\label{fig:aevstlong}
  	}
\end{figure}
\subsection{Phase-space structure investigation}
\label{sec:invest}
To investigate the long-term phase space structure further, we now consider $1001$ initial conditions.  In each, the $x$-position of the test particle has been perturbed from the fiducial initial conditions \eqref{eq:ic} according to 
\begin{equation}
\label{eq:xpert}
x_{3k} = x_3 + k 10^{-14}, 
\end{equation}
where $k$ is an integer from $0$ to $1000$.  LR is used with $h = 0.01$ and orbit information for the test particle is output every $1000$ steps, or $\Delta t = 10$ (ignoring roundoff error). Output is only generated after $t = 1500$ and stopped at $t = 3000$.  In total, there are $150150$ outputs.  These outputs are used to generate a PDF of $j$ in Fig. \ref{fig:hadden}.
\begin{figure*}
	\includegraphics[width=140mm]{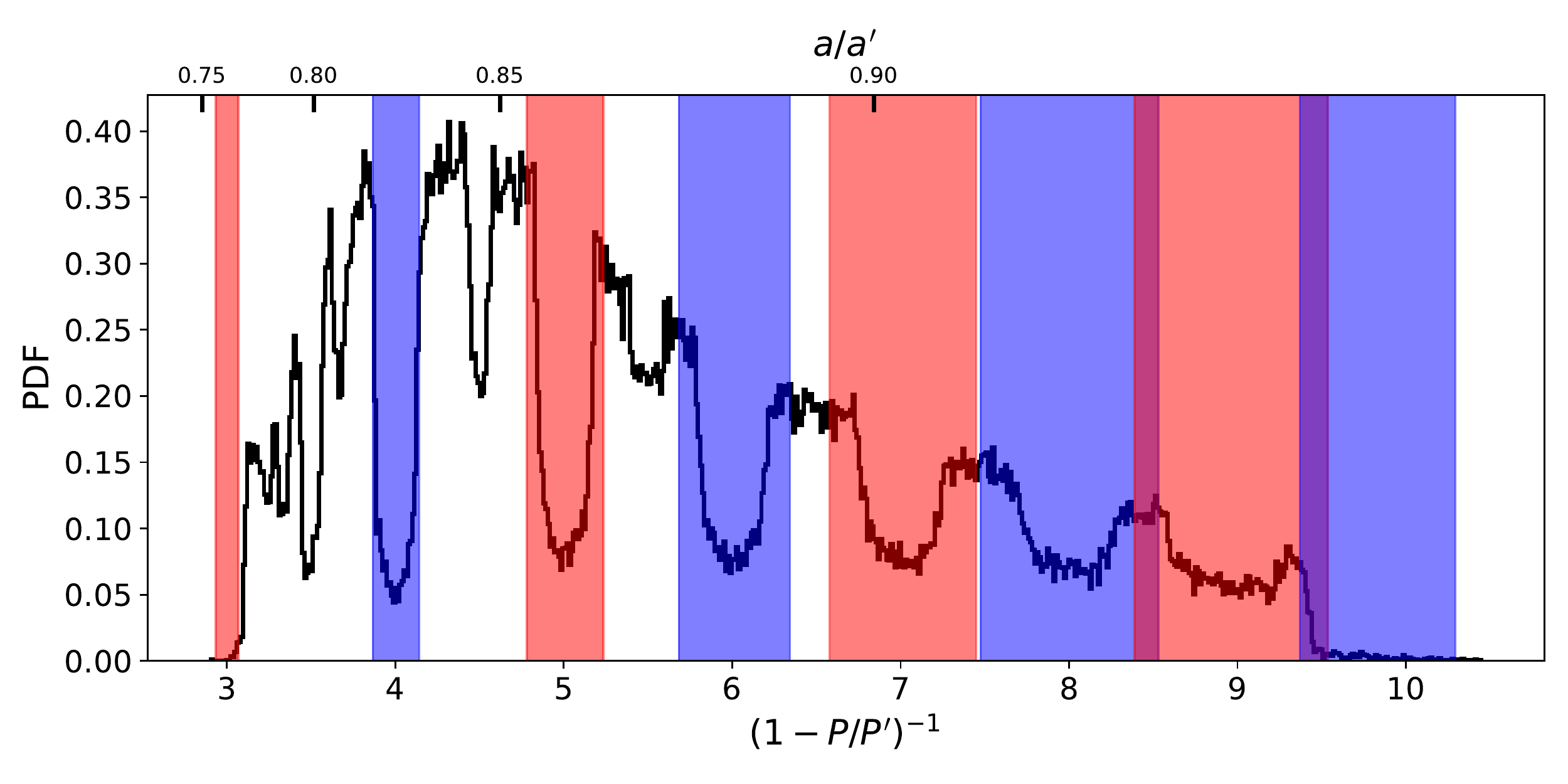}
	\caption{PDF of the semi-major axes of the test particle in $1001$ restricted three-body problems.  The initial conditions are generated according to \eqref{eq:xpert}.  They are integrated with LR, using $h = 0.01$, up to time $t = 3000$.  Output is recorded every $1000$ steps between $t = 1500$ and $t = 3000$. 
	\label{fig:hadden}
  	}
\end{figure*}
Recall the initial $j$ is 9.7.  The particle avoids integer values of $j$.  The resonance widths are indicated in red and blue.  Two colors are used to distinguish adjacent widths.  They are calculated from the intersection of the yellow separtrices in Fig. \ref{fig:hadden2} with the Jacobi constant. The agreement of the widths with the data is good except at large $j$. The disagreement at large $j$ is likely due both to more significant resonance overlap and the fact that \cite{HL18}'s predictions underestimate the widths of first-order resonances at low eccentricity.  To explore further the reason for the shape of the curve of Fig. \ref{fig:hadden}, consider a surface of section that plots $j$ versus $\lambda - \varpi$.  Output is collected at conjunction when $\lambda - \lambda^\prime$ changes sign from negative to positive.  We use WH with $h = 0.01$, run for $t = 10000$, to make this plot.  A set of $40$ initial conditions are integrated that have the same Jacobi constant, $J = -5.114872215052749$, as that of initial conditions \eqref{eq:ic}.  The result is shown in Fig. \ref{fig:sofs}.

\begin{figure*}
	\includegraphics[width=140mm]{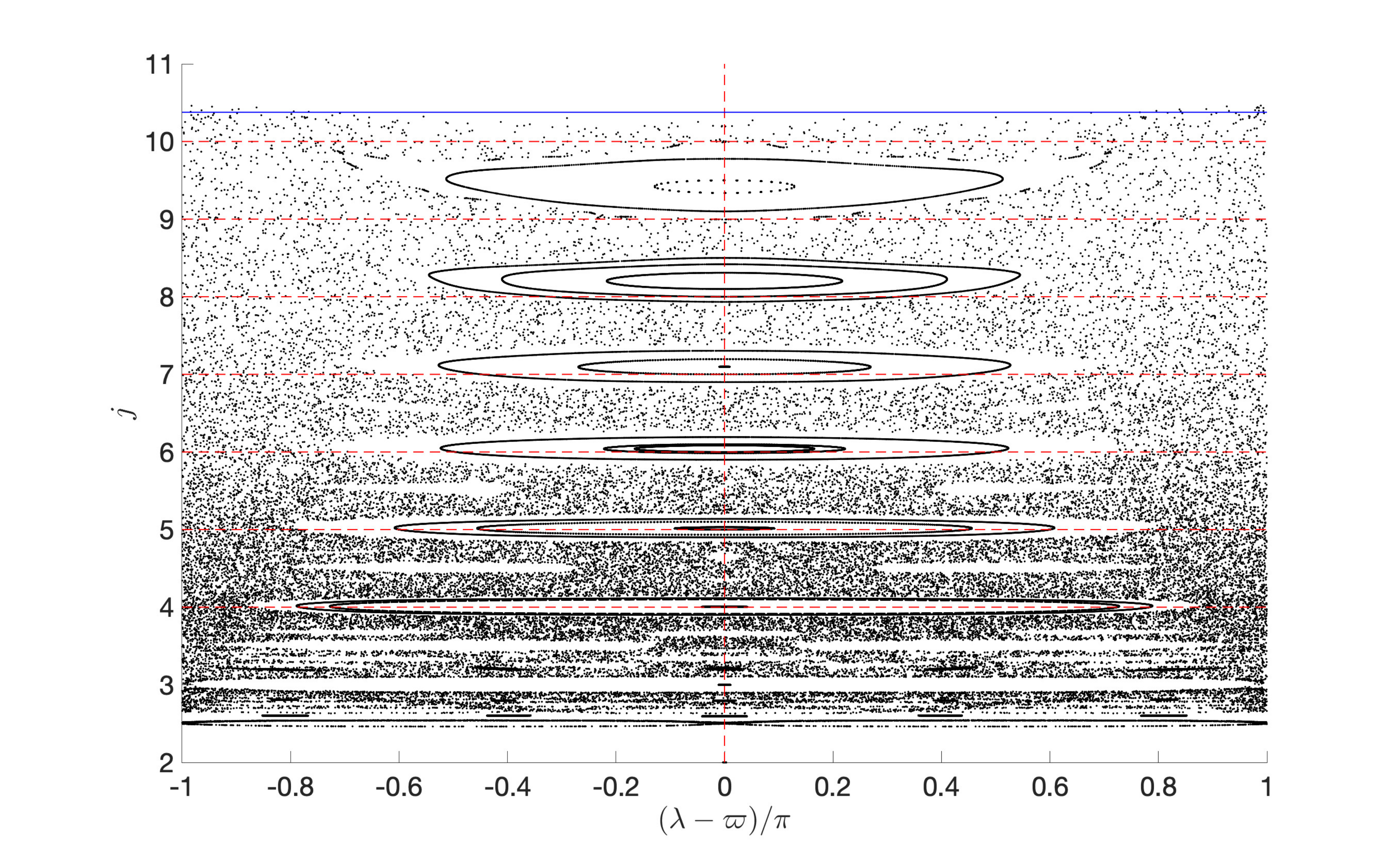}
	\caption{Surface of section plotting $j$ versus $\lambda - \varpi$.  Output is recorded at conjunction when $\lambda - \lambda^\prime$ changes sign from negative to positive.  Chaotic orbits are forbidden from accessing regular regions near first order resonances, explaining the shape of the curve of Fig. \ref{fig:hadden}.
	\label{fig:sofs}
	}
\end{figure*}
Near the first-order resonances there are forbidden areas for chaotic orbits, in agreeement with Fig. \ref{fig:hadden}. The resonance centers in Figure \ref{fig:sofs} are shifted interior to the nominal resonant locations at large $j$ both because the surface of section records osculating, rather than mean, orbital elements and because stronger perihelion precession at lower eccentricities and closer spacings shifts the semi-major axes on which the resonances are centered.  The widths of the regular regions increase with $j$.  Under $j = 3$, the phase space is regular, explaining the drop in PDF of Fig. \ref{fig:hadden}.  A blue line indicates the value of $j$ for a circular orbit.  With the surface of section plot, we can also see that PDF dips between $j = 3$, $4$, and $5$, are due to higher order MMRs. 

\section{Comparisons}
\label{sec:comp}
In Section \ref{sec:invest}, we found that $1001$ long-term integrations, using LR with $h = 0.01$ gave a PDF in semi-major axis consistent with the resonant widths in \cite{HL18}.  We now perform experiments to compare the statistics of different integrators used with different stepsizes.  $10$ sets of experiments are integrated as in Section \ref{sec:invest}.  In the first three experiments, WH is run using $h = 0.01$, $h = 0.05$, and $h = 0.15$.  In the next three, WHc is run with $h = 0.01$, $h = 0.05$, and $h = 0.15$.  Next, LR is run with $h = 0.01$, $h = 0.05$, and $h = 0.2$.  Finally, an RK experiment is done, using absolute and relative tolerances of $10^{-9}$.  But variations in the tolerance did not affect the results significantly.  {The RK method requires time derivatives of the phase space coordinates, which we obtain from the Hamiltonian set of ODEs.}  Again, as in \ref{sec:invest}, output is recorded between $t = 1500$ and $t = 3000$.  For the $h = 0.01$ , $h = 0.05$, $h = 0.15$, and $h = 0.2$ experiments, output is recorded every $1000$, $200$, $67$, and $50$ steps, respectively.  

Not all $1001$ runs in the symplectic experiments were completed, due to a failure in the Kepler advancer converging.  This is usually a sign that the timestep was too large with respect to the osculating orbital timescale, which is the period for elliptic orbits.  See also, Fig. 1, bottom panel, of \cite{WH15}.  We used the Kepler advancer of \cite{WH15}.  A list of the discarded runs is shown in Table \ref{tab:fail} , labelled by $k$ from eq. \eqref{eq:xpert}.  Experiments using larger timesteps led to a larger proportion of failed runs.  No RK runs failed; note that the RK method does not use a Kepler solver.
\begin{center}
\begin{table*}
\caption{List of runs which were not completed due to a failure of the Kepler solver converging.  $k$ labels the set of initial conditions, according to \eqref{eq:xpert}.}
\centering
\begin{tabular}{| c | c | c | c |}
	\hline
	  Experiment & WH $h = 0.15$ & WHc $h = 0.15$ & LR $h = 0.2$   \\ [3ex] \hline
	 $k$ failures & $48$, $309$, $682$, $709$, $981$  &  $374$, $463$, $526$ &  $181$, $426$, $563$  \\ \hline
\end{tabular}
\label{tab:fail}
\end{table*}
\end{center}   
\begin{center}
\begin{table*}
\caption{List of typical compute times, in code units, for running a set of initial conditions for a given method.}
\centering
\begin{tabular}{| c | c | c | c | c | c | c | c |}
	\hline
	  & $h = 0.01$ LR & $h = 0.05$ LR & $h = 0.2$ LR & $h = 0.01$ WH & $h = 0.05$ WH & $h = 0.15$ WH & RK  \\ [3ex] \hline
	 $t_{\mathrm{cpu}}$ &  50  &  10 &  2.3 &  23 &  4.7 &  2.0 &  32  \\ \hline
\end{tabular}
\label{tab:times}
\end{table*}
\end{center}   
The compute times of typical runs in each experiment are shown in Table \ref{tab:times}.  The time is in code units.  The WHc times are nearly identical to the WH times.  LR runs are about twice as expensive as WH runs as expected.  The RK run is expensive, but not the most expensive method.  In the symplectic methods, the Kepler advancer from maps of form $\mathcal A(a)$ dominate the compute time over the $\mathcal B(b)$ maps by a ratio of approximately $1.54$.  We mention that \cite{Reinetal2019b} compare the compute effort of symplectic integrators in their Table 1 using as criteria the force evaluations in $\mathcal B(b)$.  They state, `We list the theoretical cost of each method... It corresponds to the number of force evaluations per timestep and assumes all other operations take no time.''  Such an assumption will become valid only if the number of planets is large enough.

We calculate the Jacobi constant error at $t \approx 3000$ for all $1001$ runs in each experiment and compare the experiments.  We consider the final Jacobi constant error as a more reliable measure of error than the maximum error during the run or the RMS error.  The maximum error may just reflect an oscillation in error, while RMS errors may not capture secular trends.  We create PDFs of these errors for each experiment.  We plot them in Fig. \ref{fig:djpdf}, where we have separated the PDFs into three subplots according, roughly, to similar error magnitudes.
\begin{figure*}
	\includegraphics[width=140mm]{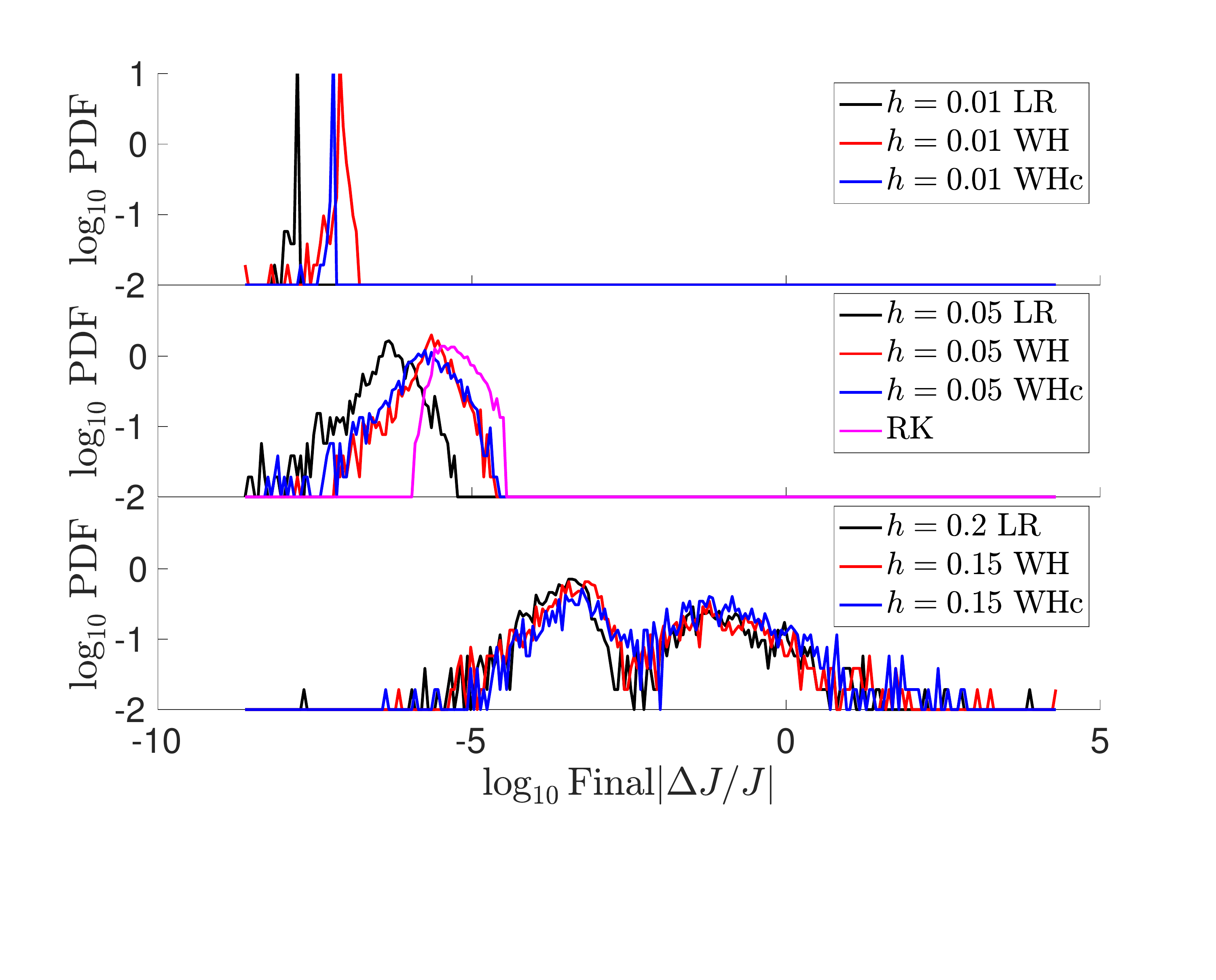}
	\caption{PDFs of the Jacobi constant error, at time $3000$, for different integrators.  The PDFs are divided into three subplots, roughly, by magnitude of the error.
	\label{fig:djpdf}
  	}
\end{figure*}
A $\log$ PDF is measured in the $y$-axis, so $0$ values have been set to $10^{-2}$.  The peaks of the PDFs of the symplectic methods scale as $h^2$ from the first to second subpanel, but the scaling breaks down from the 2nd to the 3rd subpanel, indicating uncontrolled error.  The RK experiment error is comparable to those of the experiments for $h = 0.05$.  In the third subpanel, the data have two peaks.  We will study runs from isolated peaks below.  

In Fig. \ref{fig:apdf}, we construct PDFs of the semi-major axes for all $10$ experiments, similar to those of Fig. \ref{fig:hadden}, but the $x$-axis measure has been changed to $a/a^\prime$.  Again, as in Fig. \ref{fig:djpdf}, we divide the experiments into three panels, based on approximate Jacobi constant error, except that in the bottom two subpanels, the curve for $h = 0.01$ with LR has been added for reference.  
\begin{figure*}
	\includegraphics[width=140mm]{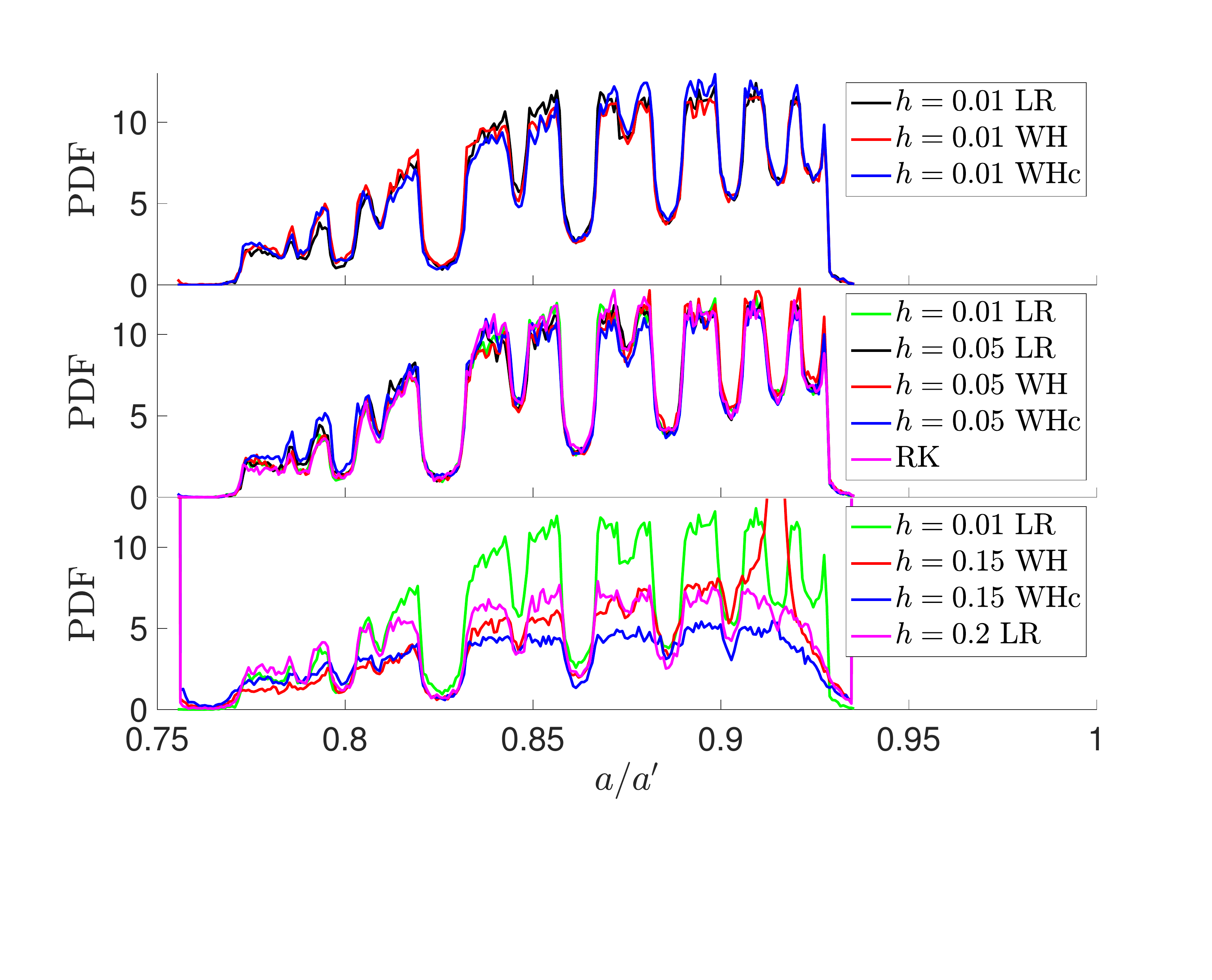}
	\caption{PDFs of the semi-major axis in units of the planetary semi-major axis.  The PDFs have been divided into three subpanels, as in Fig. \ref{fig:djpdf}, except that the $h = 0.01$ LR curve has been added to the bottom two subplots.  While the methods in the top two panels generally agree, the PDFs of the bottom panel are significantly distorted.  The methods become distorted at similar timesteps.
	\label{fig:apdf}
  	}
\end{figure*}
Recall the initial $a/a^\prime$ from Section \ref{sec:choosing}, is $a/a^\prime \approx 0.93$, which is close to the rightmost edge of the PDF.  The maximum and minimum $a/a^\prime$ in the third subpanel in the PDFs has been set as $0.9354$ and $0.7547$, respectively, and data with more extreme values has been placed into these bins.  This explains the jump in PDFs at the edges.

The PDFs are robust to error and methods used to calculate them in the top two subpanels.  Even the conventional integrator produces accurate PDFs.  To see this more quantitatively, we investigate whether differences among the PDFs can be attributed to Poisson fluctuations.  Define $\delta$ as the absolute value of the difference as a function of $a$ between a PDF curve and the curve for $h = 0.01$ LR.  {$\delta$ is a function of the number of initial conditions used to calculate statistics.}  $\overline{\delta}$ is its median.  $\overline{\delta}$ for various experiments are shown in Table \ref{tab:poisson}.  {We calculate $\overline{\delta}$ for various amounts of runs, but Figures \ref{fig:djpdf}, \ref{fig:apdf}, \ref{fig:diff}, \ref{fig:epdf}, and \ref{fig:emax} use data from $1001$ initial conditions.  New initial conditions are constructed according to eq. \eqref{eq:xpert}, and we checked such perturbations change the Jacobi constant at most by $1.2 \times 10^{-11}$.  We discarded the $k = 2530$ LR run due to a failure unrelated to the code.}
\begin{center}
\begin{table*}
\caption{Difference in PDFs from the reference PDF, from $h = 0.01$ used with LR, quantified by $\overline{\delta}$, as defined in the text.  As $h$ decreases, there is no improvement in $\overline{\delta}$.  All methods generally agree{.  WHc with $h = 0.05$ has differing statistics as the run number is increased.}  $\overline{\delta}$ decreases as the amount of runs increases, and is generally well-described by Poisson statistics.}
\centering
\begin{tabular}{| c || c| c| c| c|}
	\hline
	 integration & $\overline{\delta}$ for $10$ runs & $\overline{\delta}$ for $100$ runs & $\overline{\delta}$ for $1001$ runs & $\overline{\delta}$ for $10001$ runs \\ [3ex] \hline
	 $h = 0.05$ LR & $2.7$ & $0.88$ & $0.26$ $$ & $0.077$     \\ \hline
	 $h = 0.05$ WH & $2.7$ & $0.78$ & $0.27$ $$ & $0.13$  \\ \hline
	 $h = 0.05$ WHc & $2.7$ & $0.88$ & $0.39$ $$ &$0.41$ \\ \hline
	 RK & $2.7$ & $0.97$ & $0.20$ $$ & 0.094 \\ \hline
	 $h = 0.01$ WH & $2.4$ & $0.82$ & $0.28$ $$ & $0.072$  \\ \hline
	 $h = 0.01$ WHc & $1.8$ & $0.88$ & $0.35$ $$ & $0.11$  \\ \hline
\end{tabular}
\label{tab:poisson}
\end{table*}
\end{center}
We see in Table \ref{tab:poisson} that even though some runs give smaller Jacobi constant error, it does not improve $\overline{\delta}$.  $\overline{\delta}$ usually decreases as the number of runs increases.  This is explained by Poisson statistics, which predicts roughly that as the number of runs increases by $b$, $\overline{\delta}$ decreases as $1/\sqrt{b}$.  {However, Poisson statistics fail to explain $\overline{\delta}$ for WHc $h = 0.05$, which does not change as the number of runs goes from $1001$ to $10001$.  Applying the first correction in this case gives a Jacobi constant error of $-3.2 \times 10^{-6}$, and it may be that WHc is therefore computing statistics corresponding to different Jacobi constants.  It is surprising that the median absolute error for WH $h = 0.05$ runs is $2.3 \times 10^{-6}$ while the median error for WHc $h = 0.05$ runs is a smaller $1.6 \times 10^{-6}$.  Thus, we have found a case when a smaller Jacobi constant error actually gives phase space statistics which are unreliable.}  These results indicate that improving an integrator did not yield better phase space statistics, increasing the amount of initial conditions used did.  {It should be expected that if the amount of initial conditions increases further, systematic PDF differences would arise.  According to our work, these differences would be exceedingly small, smaller than the limits in Table \ref{tab:poisson}.}

Once the methods lose their error scaling in the third subpanel, the PDFs become clearly distorted and the methods become unreliable.  The most distorted PDF is arguably the WHc $h = 0.15$ one, according to $\overline{\delta}$.  There is a { peak in the WH $h = 0.15$ PDF centered to two significant figures at $j = 8$ ($a/a^\prime \approx 0.91$)}.  This spike in the PDF comes {uniformly} from data in the $1001$ runs and is sensitive to integration parameters.  Reruning this experiment using $h = 0.149$ instead, with WH, moves the peak to $j = 7$ nearly exactly.  It is unclear whether this peak is due to a timestep resonance effect \citep{RH99} or a feature of the real system with a different Jacobi constant.  We have only observed such peaks in additional experiments when the PDF is already significantly distorted otherwise.  We explored whether the PDFs in the third subpanel improved if we only included runs with Jacobi constant error smaller than some threshold.  The PDFs remained distorted.  We also analyzed the WH $h = 0.15$ PDF only using runs with Jacobi constant error approximately in the first peak of the error PDF of Fig. \ref{fig:djpdf}--- i.e., runs satisfying $-2.3 < \log_{10} |\Delta J/J| < 0.76$, but it caused the semi-major axis PDF to become even more distorted.  {In one test, we were able to improve the PDFs of the third subpanel.  We removed any integrations for which $0.76 < a/a^\prime < 0.94$ did not hold for all times.  This led to $\overline{\delta} = 1.10$ for WHc, which is still far worse than the $\overline{\delta}$ in the top two subpanels.
}

Past a critical timestep, a method's PDF becomes increasingly distorted, according to $\overline{\delta}$.  For WH, this critical timestep is roughly $h = 0.05$.  $\overline{\delta}$ grows as $0.27$, $0.54$, and $1.07$ for $h = 0.5$, $h = 0.13$ (not shown in Fig. \ref{fig:apdf}), and $h = 0.15$, respectively.  We also explore $h = 0.10$ (not shown in the PDF) in a surface of section in Fig. \ref{fig:sofs10}.  Fewer than the $40$ orbits of Fig. \ref{fig:sofs} are integrated because orbits were discarded when the Kepler solver failed to converge.  The integration time was $t = 5000$.  The area of the chaotic sea is now larger, due to chaos from overlapping step-size resonances, making more of phase space accessible to the test-particle.

\begin{figure*}
	\includegraphics[width=140mm]{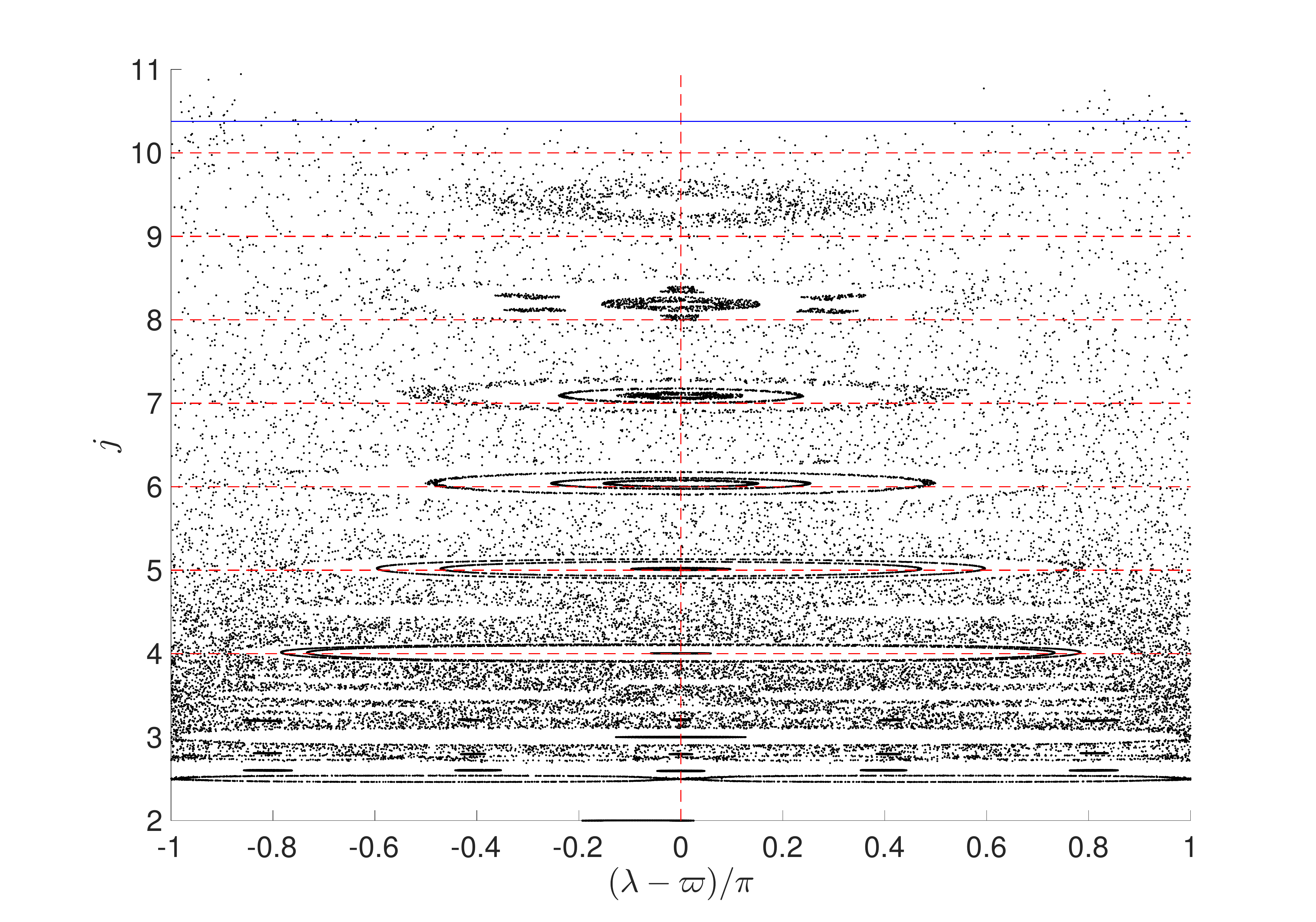}
	\caption{Surface of section plotting $j$ versus $\lambda - \varpi$, using WH with $h = 0.10$.  The integrations are run for $t = 5000$.  Compared to Fig. \ref{fig:sofs}, the area of the chaotic sea has increased.
	\label{fig:sofs10}
	}
\end{figure*}

The fact that the semi-major axis statistics are robust to different methods and errors is in broad agreement with the democratic $N$-body experiments discussed in Sec. \ref{sec:lit}.  Higher order symplectic methods provide no clear advantages.  The symplectic methods, regardless of order, failed at approximately the same stepsize.  LR allows larger steps as seen in the third subplot.  Its PDF is less distorted.  Note that its intermediate steps from eq. \eqref{eq:LR} are approximately half the magnitude of those of WH, so it's reasonable to believe LR can resolve more timescales than WH for a given $h$.  {The LR effective timestep, using \cite{Wisdom2018} criteria, is $h_{\mathrm{eff}} \approx h/2$.}  {Finally, we find that if we introduce a small inclination to the test particle of $2$ degrees, we already see significant differences among the PDFs for $h = 0.05$.}

Although we find consistency in the PDFs, it's possible that as the amount of initial conditions becomes larger, the PDFs of the top two panels of Fig. \ref{fig:apdf} will converge to different shapes, not explained by Poisson statistics.  However, we see that already for large number of initial conditions and methods, the statistics agree.

\subsection{Resolution of pericenter}
\label{eq:resol}
The effective period at pericenter for the test particle is \citep{W15}, {
\begin{equation}
\tau_{\dot{f}} = \frac{2 \pi}{\dot{f}_{\mathrm{max}}},
\end{equation}
where $\dot{f}_{\mathrm{max}}$ is the maximum time derivative of the true anomaly.  At time $0$,} {
\begin{equation}
\tau_{\dot{f}} = 2 \pi \sqrt{\frac{(1-e)^3}{1+e} a^3} \approx 0.82.
\end{equation} 
}Roughly, the largest stepsize producing an accurate WH PDF is $h = 0.05$, as noted above, yielding $16$ steps per $\tau_{\dot{f}}$.  According to \cite{W15}, this is approximately the minimum amount of steps required to resolve pericenter.  Note, however, that we use different coordinates from \cite{W15}, who uses Jacobi coordinates.  Also, our figure of $16$ steps per $\tau_{\dot{f}}$ should be considered an upper bound since $\tau_{\dot{f}}$ can decrease over the integration.  Our results roughly suggest that if pericenter is resolved, the statistics are valid, but this connection should be explored in more detail.

\subsection{Diffusion in the Jacobi constant}
\label{sec:diff}
It is well known that the energy error of symplectic maps is bounded, as,
\begin{equation}
\label{eq:errb}
    |H(t)-H(0)| = \mathcal O(e^{-h_0/h}) \mathrm{~for~} t < e^{h_0/h},
\end{equation}
where $h_0$ is is positive \citep{hair06,Engleetal2005,BenettinGiorgilli1994}.  {In a rotating frame, the Jacobi integral is the time evolution operator of the restricted three-body problem, and thus, bounds \eqref{eq:errb} apply to it.}  Define a variance in the Jacobi constant,
\begin{equation}
\label{eq:diff}
\sigma^2(t) = \frac{\langle \left( J(t) - J_0 \right)^2 \rangle}{J_0^2} = (t/t_{\mathrm{sc}})^n + a_1,
\end{equation}
with $a_1$ a constant.  An expectation value is calculated by averaging over all initial conditions.  $J_0$ is the initial Jacobi constant.  In a diffusive process, $n=1$ \citep{Einstein1905}.  In Fig. \ref{fig:diff}, we plot $\sigma^2(t)$ for different integrators.  For the RK method, $n = 3$  For the $h = 0.05$ symplectic methods, the drift is {approximately} described by a diffusive process with $n = 1$.  {We have confirmed similar behavior in other symplectic maps.}  For small $h$, $n = 0$: there is no sign of diffusion.  In experiments of other symplectic maps, we have found diffusion can appear at later times.  The lack of energy drift with small $h$ confirms we are not seeing roundoff error effects.  {Because the} PDFs become distorted when $\sigma^2(t)$ {becomes large, it's clear that eventually the statistics of both symplectic methods and the RK method will be unreliable.} 

\begin{figure*}
	\includegraphics[width=140mm]{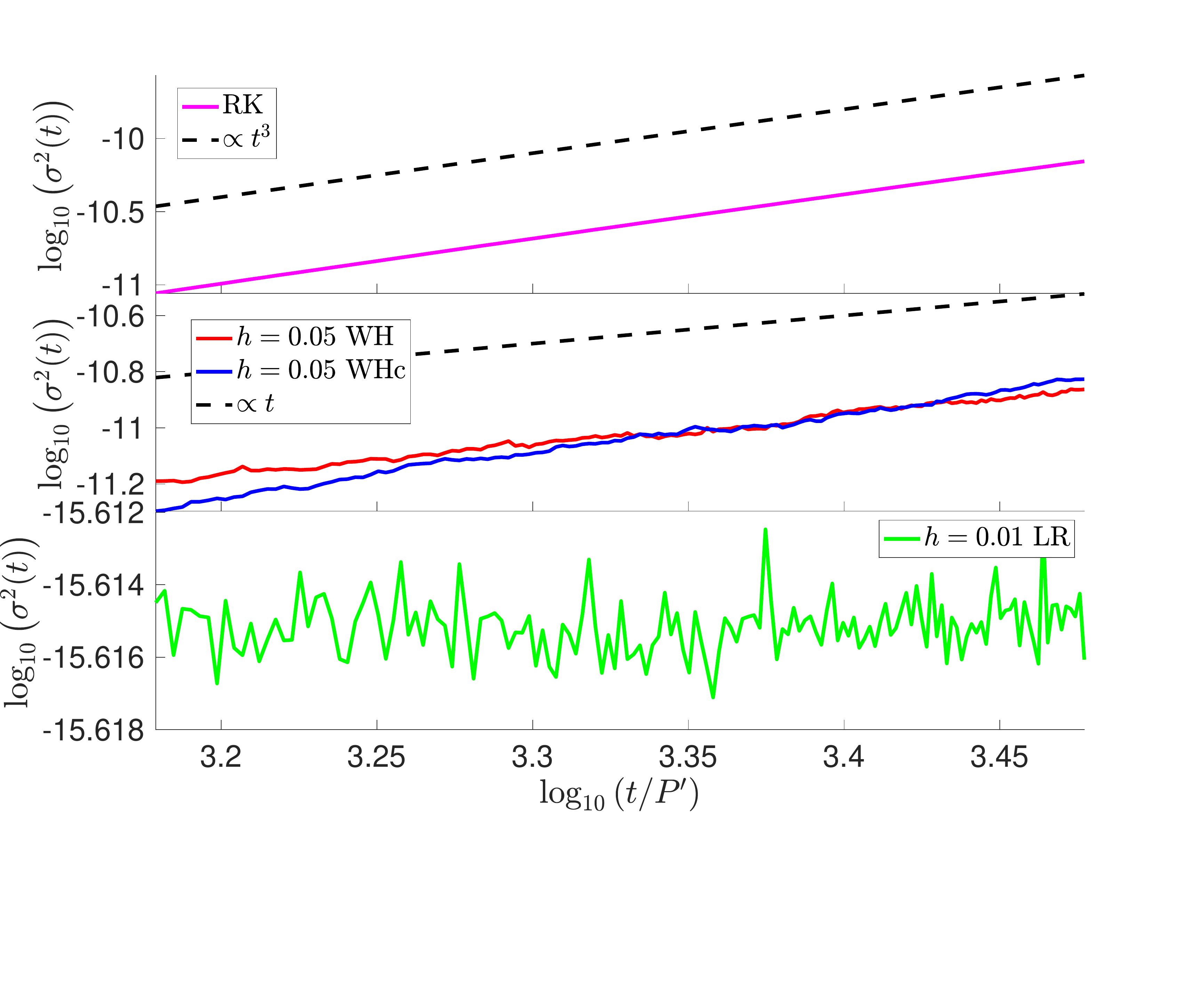}
	\caption{Growth in $\sigma^2(t)$ for some integrations.  For RK, $\sigma^2(t) \propto t^3$, while for the $h = 0.05$ methods, $\sigma^2(t) \propto t$ roughly.  There is no clear dependence on $t$ for $h$ small.
  	}
	\label{fig:diff}
\end{figure*}

{The WH $h = 0.05$ data from the second panel of Fig. \ref{fig:diff} is explored more closely in Fig.  \ref{fig:jacob}.  The error in the Jacobi constant is plotted as a function of time for the $1000$ initial conditions.  Note the range of times has been extended to $t \in (0,6000)$.  At $t \lesssim 200$, the $1000$ curves are indistinguishable on the chosen scale.  For $t \gtrsim 1500$, the average of the absolute error grows as $\sqrt{t}$, but for $t \lesssim 1500$, the average does not grow significantly in time.  Also at $t \lesssim 1500$, no diffusion in $\sigma^2(t)$ is present. }

\begin{figure}
	\includegraphics[width=90mm]{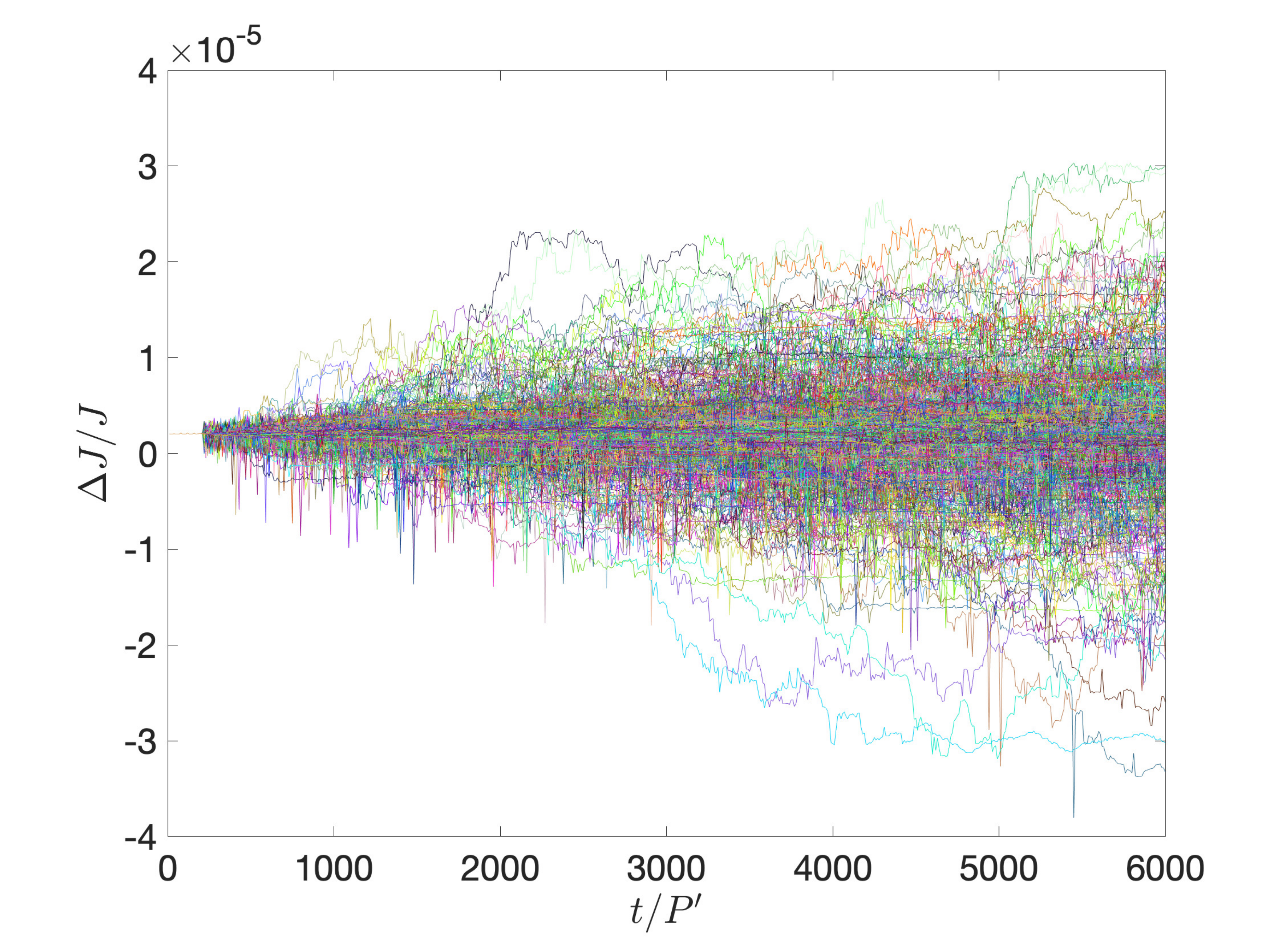}
	\caption{{Error in the Jacobi constant for the $1000$ initial conditions.  WH with $h = 0.05$ is used.}
  	}
	\label{fig:jacob}
\end{figure}

\subsection{Additional statistics}
We construct two more statistics.  Fig. \ref{fig:epdf} plots the PDFs of the osculating eccentricities for the $10$ experiments.
\begin{figure*}
	\includegraphics[width=140mm]{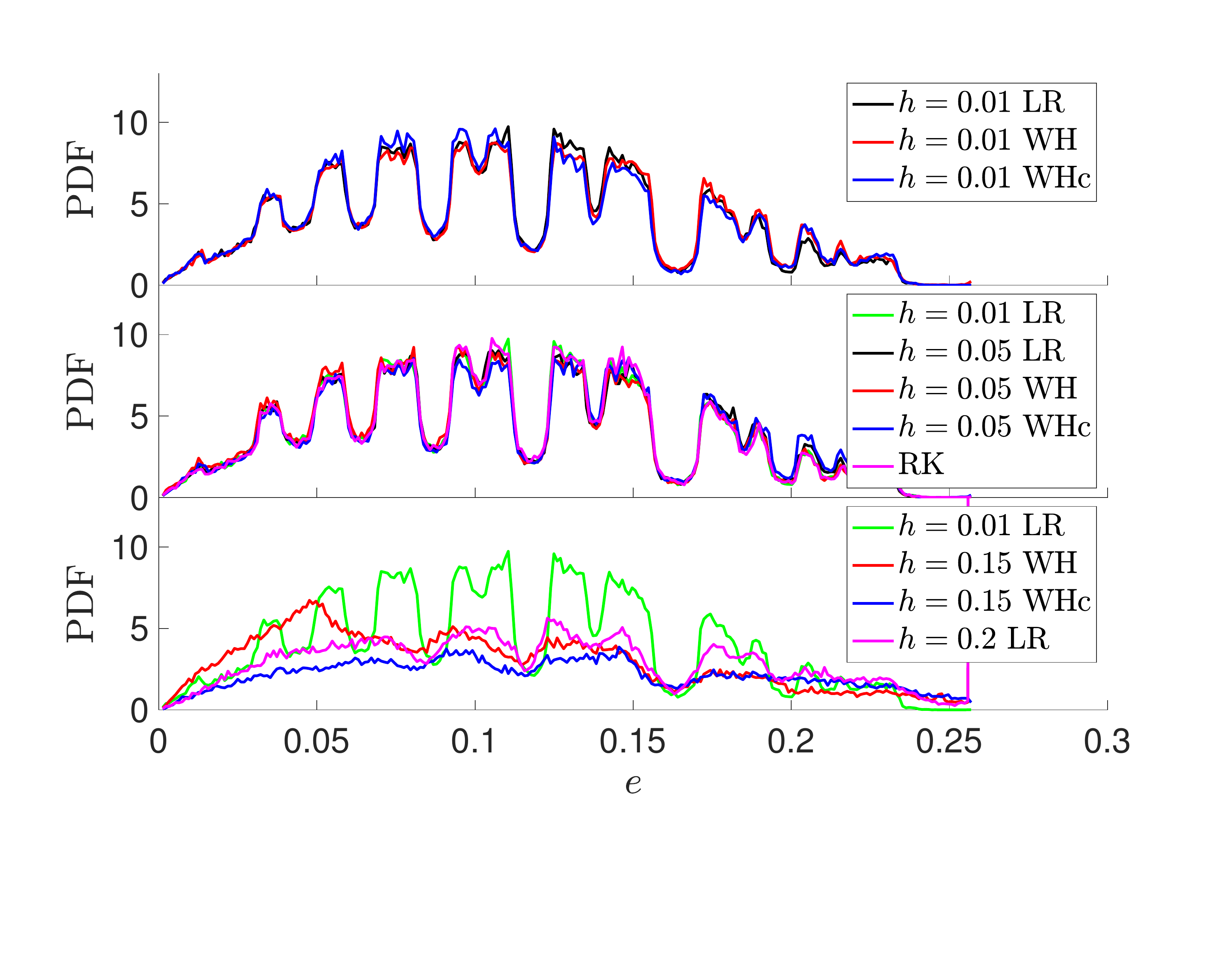}
	\caption{PDFs of the osculating test-particle eccentricities, as in Fig. \ref{fig:apdf}.
	\label{fig:epdf}
  	}
\end{figure*}
{Due to the existence of the Jacobi constant and symmetry in the angle variables, statistics of the planar restricted three-body problem can be described by $a$ or $e$.  If the PDF for $a$ and $e$ is $P_a(a)$ and $P_e(a)$, respectively, then $P_a (a) da = P_e (e) de$.  $da/de$ is found approximately from the Jacobi constant in form,
\begin{equation}
\label{eq:jacobi}
J = \frac{m_1}{2a} \frac{1+e}{1-e} - \Omega \sqrt{m_1 a (1-e^2)} - \frac{m_1}{ a(1-e) - x_{1}} - \frac{m_2}{a (1-e) - x_{2}},
\end{equation}
where $\Omega = 2 \pi$, $x_1$ is the initial $x$-coordinate of $m_1$ and $x_2$ is the $x$-coordinate of $m_2$.  We checked that converting the PDF of Fig. \ref{fig:apdf} to an $e$ PDF through eq. \eqref{eq:jacobi} matches Fig. \ref{fig:epdf}, except at small eccentricities.}

The experiments have been divided into three subplots as in Fig. \ref{fig:apdf}.  In the bottom subplot, maximum and minimum eccentricity bins have been set as $0.2569$ and $2.2552 \times 10^{-4}$, respectively.  Recall from Section \ref{sec:choosing} that the initial eccentricity is $\approx 0.036$.

The results of the eccentricity PDFs agree with those of the $a$ PDFs.  The statistics are reliable for all methods until the errors become relatively large in the third subplot.  The significant peak in Fig. \ref{fig:apdf} is gone in Fig. \ref{fig:epdf}.  Many runs used for the third subplot of Fig. \ref{fig:epdf} achieved hyperbolic eccentricities.  For the $h = 0.02$ LR, $h = 0.015$ WH, and $h = 0.015$ WHc curves, 122, 114, and $193$ runs, respectively, had hyperbolic orbits.

Finally, we compute a cumulative distribution function of the maximum eccentricity, $e_{\mathrm{max}}$ in each run in Fig. \ref{fig:emax}.
\begin{figure*}
	\includegraphics[width=140mm]{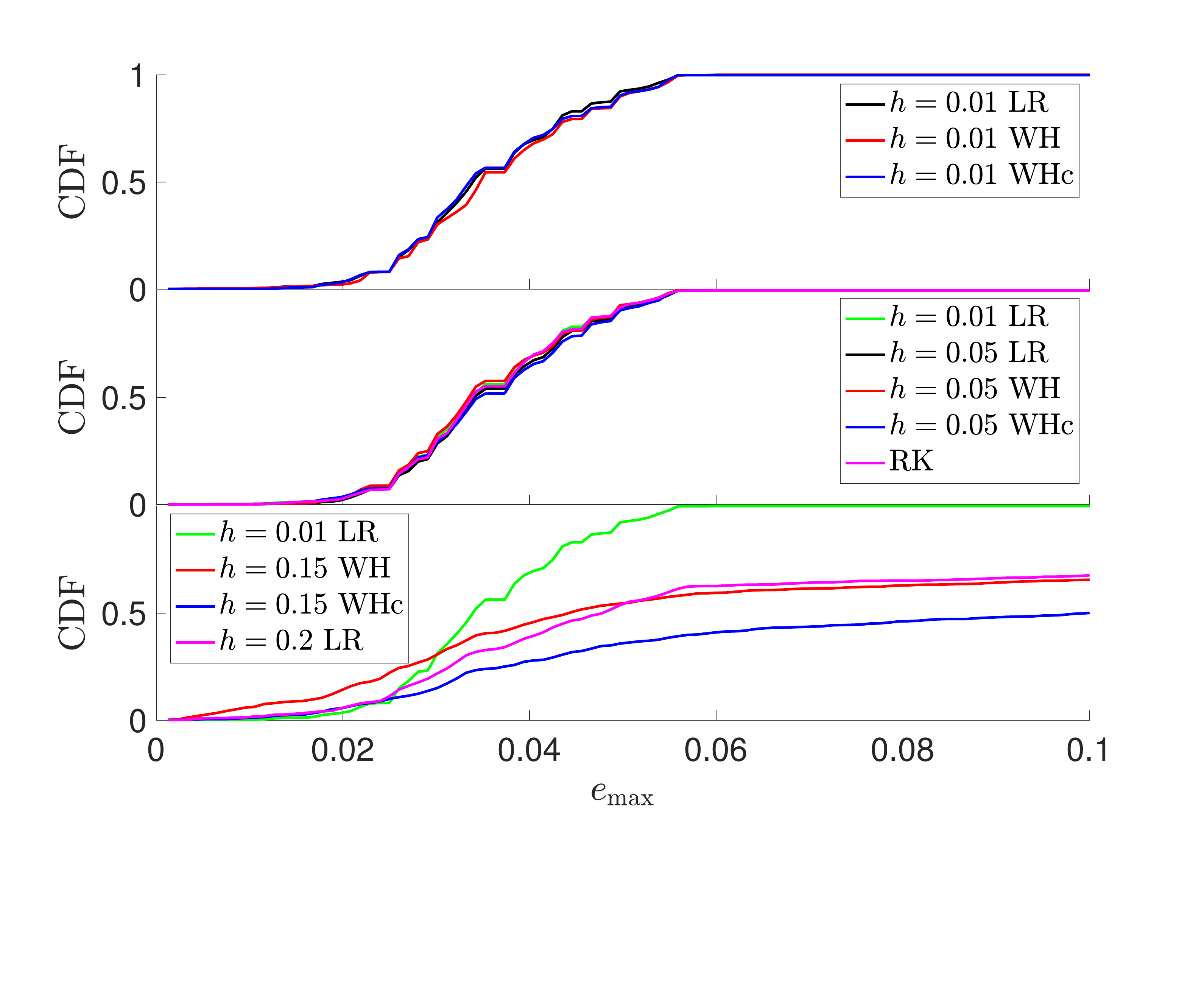}
	\caption{Cumulative distribution of the maximum eccentricity attained for a given set of initial conditions and integrator.  The distributions are subdivided into three subplots, as in Fig. \ref{fig:djpdf}.
	\label{fig:emax}
  	}
\end{figure*}
So each experiment has $1001$ $e_{\mathrm{max}}$ values that we bin.  A similar statistic is seen in \cite[Fig. 1]{Laskar08} for the Solar System.  The $e_{\mathrm{max}}$ CDFs show broad agreements until the third subpanel, once again.  The maximum CDF difference in the top two panels again comes from the $h = 0.05$ WHc run.
{
\section{Discussion}
\subsection{On the {use} of higher order symplectic integrators}
\label{sec:impact}

{
For long-term solutions, using low-order symplectic integrators, or sometimes conventional integrators, appears optimal.  For short-term solutions, there are various cases.  If the number of steps is small, a conventional integrator is sufficient.  If the number of steps is large and only qualitative information is needed, one might use a low-order symplectic integrator.  Finally, if the number of steps is large, and angle variable information is required, a higher-order symplectic method or conventional integrator might be optimal, depending on the accuracy required.}

While we have studied simple two degree of freedom systems, it's worth discussing what would happen with higher dimensional systems like the Solar System.  In this case, computing an energy error may not give us a good handle of the reliability of an integration, but, as discussed in Section \ref{eq:resol}, resolving pericenter in the orbits could be a good criteria for reliability.  The reliability of Solar System integrations could be addressed with more computing resources.  Our arguments need modification if working in extended precision. }
\subsection{Conclusions}
\label{sec:conc}
We have studied the reliability of long-term $N$-body integrations by comparing the phase-space structure of the restricted three-body problem calculated in \cite{HL18} to phase-space statistics of large number of $N$-body integrations.  By long-term, we refer to enough Lyapunov times such that memory of the initial conditions has vanished.  {This is perhaps not the conventional definition of long-term.}  Errors introduced by finite precision arithmetic at the $10^{-16}$ level, errors in the initial conditions, or errors in the integration method itself cause the $N$-body problem the computer solves to be completely distinct from the problem corresponding to the initial conditions after at most $\approx 36$ Lyapunov times (when working with double precision arithmetic).  In our integrations, we recorded data between $100$ and $200$ Lyapunov times after the start time.  We computed PDFs of the semi-major axis and eccentricity, finding accurate integrations avoided resonances.  As compared to a conventional Runge--Kutta method, there was no clear advantage in accuracy to using a symplectic integrator or the high-order symplectic integrators of \cite{Butcher69,WHT96,W06,LR01}.  This argument holds for higher order symplectic corrector methods and other high order symplectic integrators \citep{WHT96,W06,LR01,Reinetal2019b} because they are composed of the same maps $\mathcal A(a)$ and $\mathcal B(b)$ of Section \ref{sec:methods}, even if they have modified kicks. The advantage of a symplectic integrator over a conventional integrator is in the speed, and this holds only if the symplectic integrator is low order and (relatively) low-accuracy in error of first integrals like the Jacobi constant. Such an advantage can get ruined if symplecticity is broken at a few points in phase space \citep{H19}.  Once the timestep was made large enough that the error in the Jacobi constant became uncontrolled, all symplectic integrators failed to give correct statistics, regardless of their error or accuracy.  This should not be surprising since the symplectic methods are composed of the same basic maps that are functions of the timestep.  If one method has timestep too large to resolve the orbital dynamics, the others will suffer the same issue, regardless of order.  The situation is slightly different with the \cite{LR01} LR method, whose operators are functions of parameters at most $0.58$ times the magnitude of the timestep.  Thus, we have found evidence that the PDFs for LR remain accurate to slightly larger timesteps than the other methods.  But LR is expensive to run.

One surprising result of our work is that a Jacobi constant error of at most about $5 \times 10^{-6}$ (the median error of the RK runs) was allowed for accurate statistics to be reproduced.  In other studies of democratic $N$-body problems, the error was allowed to be about $4$ orders of magnitude higher.

We also challenge the idea that symplectic methods are preferable for near-integrable Hamiltonians \citep{chan90} for long-term studies.  {The conventional Runge--Kutta integrator we used has significant truncation error, and over even longer timescales than those we studied this error might become important.  However, it is simple to develop a conventional method where the truncation error is kept below machine precision so that the Runge--Kutta problems are avoided.  These conventional methods would remain reliable longer than symplectic methods, which are subject to some truncation error (Section \ref{sec:diff}).  Of course, such methods might be slower than low order symplectic integrators.}  This study supports the notion that the best way to explore long-term $N$-body orbital dynamics is to run several integrations using low-order, `low-accuracy' symplectic integrations, and calculating their statistics.  
\section{Acknowledgements}
\label{sec:ack}
{We thank Dan Tamayo and Carl Rodriguez for suggestions, Robert McLachlan for discussions, and Antoine Petit for a helpful referee report.}
\bibliographystyle{mnras}
\bibliography{paper}

\begin{thebibliography}{}
\makeatletter
\relax
\def\mn@urlcharsother{\let\do\@makeother \do\$\do\&\do\#\do\^\do\_\do\%\do\~}
\def\mn@doi{\begingroup\mn@urlcharsother \@ifnextchar [ {\mn@doi@}
  {\mn@doi@[]}}
\def\mn@doi@[#1]#2{\def\@tempa{#1}\ifx\@tempa\@empty \href
  {http://dx.doi.org/#2} {doi:#2}\else \href {http://dx.doi.org/#2} {#1}\fi
  \endgroup}
\def\mn@eprint#1#2{\mn@eprint@#1:#2::\@nil}
\def\mn@eprint@arXiv#1{\href {http://arxiv.org/abs/#1} {{\tt arXiv:#1}}}
\def\mn@eprint@dblp#1{\href {http://dblp.uni-trier.de/rec/bibtex/#1.xml}
  {dblp:#1}}
\def\mn@eprint@#1:#2:#3:#4\@nil{\def\@tempa {#1}\def\@tempb {#2}\def\@tempc
  {#3}\ifx \@tempc \@empty \let \@tempc \@tempb \let \@tempb \@tempa \fi \ifx
  \@tempb \@empty \def\@tempb {arXiv}\fi \@ifundefined
  {mn@eprint@\@tempb}{\@tempb:\@tempc}{\expandafter \expandafter \csname
  mn@eprint@\@tempb\endcsname \expandafter{\@tempc}}}

\bibitem[\protect\citeauthoryear{Benettin \& Giorgilli}{Benettin \&
  Giorgilli}{1994}]{BenettinGiorgilli1994}
Benettin G.,  Giorgilli A.,  1994, Journal of Statistical Physics, 74, 1117

\bibitem[\protect\citeauthoryear{{Binney} \& {Tremaine}}{{Binney} \&
  {Tremaine}}{2008}]{BT08}
{Binney} J.,  {Tremaine} S.,  2008, {Galactic Dynamics: Second Edition}.
Princeton University Press

\bibitem[\protect\citeauthoryear{{Boekholt} \& {Portegies Zwart}}{{Boekholt} \&
  {Portegies Zwart}}{2015}]{BP15}
{Boekholt} T.,  {Portegies Zwart} S.,  2015, \mn@doi [Computational
  Astrophysics and Cosmology] {10.1186/s40668-014-0005-3}, \href
  {http://adsabs.harvard.edu/abs/2015ComAC...2....2B} {2, 2}

\bibitem[\protect\citeauthoryear{Butcher}{Butcher}{1969}]{Butcher69}
Butcher J.,  1969, in Conference on the numerical solution of differential
  equations. pp 133--139

\bibitem[\protect\citeauthoryear{{Channell} \& {Scovel}}{{Channell} \&
  {Scovel}}{1990}]{chan90}
{Channell} P.~J.,  {Scovel} C.,  1990, Nonlinearity, 3, 231

\bibitem[\protect\citeauthoryear{{Chin}}{{Chin}}{1997}]{Chin97}
{Chin} S.~A.,  1997, \mn@doi [Physics Letters A]
  {10.1016/S0375-9601(97)00003-0}, \href
  {https://ui.adsabs.harvard.edu/abs/1997PhLA..226..344C} {226, 344}

\bibitem[\protect\citeauthoryear{{Chin} \& {Chen}}{{Chin} \&
  {Chen}}{2005}]{ChinandChen05}
{Chin} S.~A.,  {Chen} C.~R.,  2005, \mn@doi [Celestial Mechanics and Dynamical
  Astronomy] {10.1007/s10569-004-4622-z}, \href
  {https://ui.adsabs.harvard.edu/abs/2005CeMDA..91..301C} {91, 301}

\bibitem[\protect\citeauthoryear{{Cincotta}, {Giordano}  \&
  {Sim{\'o}}}{{Cincotta} et~al.}{2003}]{Cincottaetal2003}
{Cincotta} P.~M.,  {Giordano} C.~M.,   {Sim{\'o}} C.,  2003, \mn@doi [Physica D
  Nonlinear Phenomena] {10.1016/S0167-2789(03)00103-9}, \href
  {https://ui.adsabs.harvard.edu/abs/2003PhyD..182..151C} {182, 151}

\bibitem[\protect\citeauthoryear{{Dehnen} \& {Hernandez}}{{Dehnen} \&
  {Hernandez}}{2017}]{DH17}
{Dehnen} W.,  {Hernandez} D.~M.,  2017, \mn@doi [\mnras]
  {10.1093/mnras/stw2758}, \href
  {http://adsabs.harvard.edu/abs/2017MNRAS.465.1201D} {465, 1201}

\bibitem[\protect\citeauthoryear{{Einstein}}{{Einstein}}{1905}]{Einstein1905}
{Einstein} A.,  1905, \mn@doi [Annalen der Physik] {10.1002/andp.19053220806},
  \href {https://ui.adsabs.harvard.edu/abs/1905AnP...322..549E} {322, 549}

\bibitem[\protect\citeauthoryear{Engle, Skeel  \& Drees}{Engle
  et~al.}{2005}]{Engleetal2005}
Engle R.~D.,  Skeel R.~D.,   Drees M.,  2005, \mn@doi [J. Comput. Phys.]
  {10.1016/j.jcp.2004.12.009}, 206, 432

\bibitem[\protect\citeauthoryear{{Farr{\'e}s}, {Laskar}, {Blanes}, {Casas},
  {Makazaga}  \& {Murua}}{{Farr{\'e}s} et~al.}{2013}]{Farresetal2013}
{Farr{\'e}s} A.,  {Laskar} J.,  {Blanes} S.,  {Casas} F.,  {Makazaga} J.,
  {Murua} A.,  2013, \mn@doi [Celestial Mechanics and Dynamical Astronomy]
  {10.1007/s10569-013-9479-6}, \href
  {https://ui.adsabs.harvard.edu/abs/2013CeMDA.116..141F} {116, 141}

\bibitem[\protect\citeauthoryear{{Goodman}, {Heggie}  \& {Hut}}{{Goodman}
  et~al.}{1993}]{GHH93}
{Goodman} J.,  {Heggie} D.~C.,   {Hut} P.,  1993, \mn@doi [\apj]
  {10.1086/173196}, \href
  {https://ui.adsabs.harvard.edu/abs/1993ApJ...415..715G} {415, 715}

\bibitem[\protect\citeauthoryear{{Hadden} \& {Lithwick}}{{Hadden} \&
  {Lithwick}}{2018}]{HL18}
{Hadden} S.,  {Lithwick} Y.,  2018, \mn@doi [\aj] {10.3847/1538-3881/aad32c},
  \href {https://ui.adsabs.harvard.edu/abs/2018AJ....156...95H} {156, 95}

\bibitem[\protect\citeauthoryear{{Hairer}, {Lubich}  \& {Wanner}}{{Hairer}
  et~al.}{2006}]{hair06}
{Hairer} E.,  {Lubich} C.,   {Wanner} G.,  2006, {Geometrical Numerical
  Integration}, 2nd edn.
Springer Verlag, Berlin

\bibitem[\protect\citeauthoryear{{Heggie}}{{Heggie}}{1991}]{Heggie91}
{Heggie} D.~C.,  1991, in {Roeser} S.,  {Bastian} U.,  eds, Predictability,
  Stability, and Chaos in N-Body Dynamical Systems. pp 47--62

\bibitem[\protect\citeauthoryear{{Heggie} \& {Hut}}{{Heggie} \&
  {Hut}}{2003}]{HH03}
{Heggie} D.,  {Hut} P.,  2003, {The Gravitational Million-Body Problem: A
  Multidisciplinary Approach to Star Cluster Dynamics}.
Cambridge University Press

\bibitem[\protect\citeauthoryear{{H{\'e}non}}{{H{\'e}non}}{1971}]{Henon1971}
{H{\'e}non} M.~H.,  1971, \mn@doi [\apss] {10.1007/BF00649201}, \href
  {https://ui.adsabs.harvard.edu/abs/1971Ap%26SS..14..151H} {14, 151}

\bibitem[\protect\citeauthoryear{{Hernandez}}{{Hernandez}}{2016}]{H16}
{Hernandez} D.~M.,  2016, \mn@doi [\mnras] {10.1093/mnras/stw569}, \href
  {http://adsabs.harvard.edu/abs/2016MNRAS.458.4285H} {458, 4285}

\bibitem[\protect\citeauthoryear{{Hernandez}}{{Hernandez}}{2019a}]{H19}
{Hernandez} D.~M.,  2019a, \mn@doi [\mnras] {10.1093/mnras/stz884}, \href
  {https://ui.adsabs.harvard.edu/abs/2019MNRAS.486.5231H} {486, 5231}

\bibitem[\protect\citeauthoryear{{Hernandez}}{{Hernandez}}{2019b}]{H19b}
{Hernandez} D.~M.,  2019b, \mn@doi [\mnras] {10.1093/mnras/stz2662}, \href
  {https://ui.adsabs.harvard.edu/abs/2019MNRAS.490.4175H} {490, 4175}

\bibitem[\protect\citeauthoryear{{Hernandez} \& {Bertschinger}}{{Hernandez} \&
  {Bertschinger}}{2015}]{HB15}
{Hernandez} D.~M.,  {Bertschinger} E.,  2015, \mn@doi [MNRAS]
  {10.1093/mnras/stv1439}, \href
  {http://adsabs.harvard.edu/abs/2015MNRAS.452.1934H} {452, 1934}

\bibitem[\protect\citeauthoryear{{Hernandez} \& {Bertschinger}}{{Hernandez} \&
  {Bertschinger}}{2018}]{HB18}
{Hernandez} D.~M.,  {Bertschinger} E.,  2018, \mn@doi [MNRAS]
  {10.1093/mnras/sty184}, \href
  {http://adsabs.harvard.edu/abs/2018MNRAS.475.5570H} {475, 5570}

\bibitem[\protect\citeauthoryear{{Hernandez} \& {Dehnen}}{{Hernandez} \&
  {Dehnen}}{2017}]{HD17}
{Hernandez} D.~M.,  {Dehnen} W.,  2017, \mn@doi [\mnras]
  {10.1093/mnras/stx547}, \href
  {http://adsabs.harvard.edu/abs/2017MNRAS.468.2614H} {468, 2614}

\bibitem[\protect\citeauthoryear{{Kinoshita}, {Yoshida}  \&
  {Nakai}}{{Kinoshita} et~al.}{1991}]{Kinoshitaetal91}
{Kinoshita} H.,  {Yoshida} H.,   {Nakai} H.,  1991, Celestial Mechanics and
  Dynamical Astronomy, \href
  {https://ui.adsabs.harvard.edu/abs/1991CeMDA..50...59K} {50, 59}

\bibitem[\protect\citeauthoryear{{Laskar}}{{Laskar}}{2008}]{Laskar08}
{Laskar} J.,  2008, \mn@doi [\icarus] {10.1016/j.icarus.2008.02.017}, \href
  {https://ui.adsabs.harvard.edu/abs/2008Icar..196....1L} {196, 1}

\bibitem[\protect\citeauthoryear{{Laskar} \& {Robutel}}{{Laskar} \&
  {Robutel}}{2001}]{LR01}
{Laskar} J.,  {Robutel} P.,  2001, Celestial Mechanics and Dynamical Astronomy,
  \href {https://ui.adsabs.harvard.edu/abs/2001CeMDA..80...39L} {80, 39}

\bibitem[\protect\citeauthoryear{{Miller}}{{Miller}}{1964}]{Miller1964}
{Miller} R.~H.,  1964, \mn@doi [\apj] {10.1086/147911}, \href
  {https://ui.adsabs.harvard.edu/abs/1964ApJ...140..250M} {140, 250}

\bibitem[\protect\citeauthoryear{{Murray} \& {Dermott}}{{Murray} \&
  {Dermott}}{1999}]{MurrayDermott99}
{Murray} C.~D.,  {Dermott} S.~F.,  1999, {Solar system dynamics}

\bibitem[\protect\citeauthoryear{{Portegies Zwart} \& {Boekholt}}{{Portegies
  Zwart} \& {Boekholt}}{2014}]{PB14}
{Portegies Zwart} S.,  {Boekholt} T.,  2014, \mn@doi [ApJ]
  {10.1088/2041-8205/785/1/L3}, 785, L3

\bibitem[\protect\citeauthoryear{{Portegies Zwart} \& {Boekholt}}{{Portegies
  Zwart} \& {Boekholt}}{2018}]{PB18}
{Portegies Zwart} S.~F.,  {Boekholt} T. C.~N.,  2018, \mn@doi [Communications
  in Nonlinear Science and Numerical Simulations]
  {10.1016/j.cnsns.2018.02.002}, \href
  {https://ui.adsabs.harvard.edu/abs/2018CNSNS..61..160P} {61, 160}

\bibitem[\protect\citeauthoryear{{Press}, {Teukolsky}, {Vetterling}  \&
  {Flannery}}{{Press} et~al.}{2002}]{press02}
{Press} W.~H.,  {Teukolsky} S.~A.,  {Vetterling} W.~T.,   {Flannery} B.~P.,
  2002, {Numerical recipes in C++ : the art of scientific computing}

\bibitem[\protect\citeauthoryear{{Quinlan} \& {Tremaine}}{{Quinlan} \&
  {Tremaine}}{1992}]{QT92}
{Quinlan} G.~D.,  {Tremaine} S.,  1992, MNRAS, \href
  {http://adsabs.harvard.edu/abs/1992MNRAS.259..505Q} {259, 505}

\bibitem[\protect\citeauthoryear{{Rauch} \& {Holman}}{{Rauch} \&
  {Holman}}{1999}]{RH99}
{Rauch} K.~P.,  {Holman} M.,  1999, \mn@doi [\aj] {10.1086/300720}, \href
  {https://ui.adsabs.harvard.edu/abs/1999AJ....117.1087R} {117, 1087}

\bibitem[\protect\citeauthoryear{{Rein} \& {Spiegel}}{{Rein} \&
  {Spiegel}}{2015}]{RS15}
{Rein} H.,  {Spiegel} D.~S.,  2015, \mn@doi [MNRAS] {10.1093/mnras/stu2164},
  \href {http://adsabs.harvard.edu/abs/2015MNRAS.446.1424R} {446, 1424}

\bibitem[\protect\citeauthoryear{{Rein} \& {Tamayo}}{{Rein} \&
  {Tamayo}}{2018}]{ReinTamayo2018}
{Rein} H.,  {Tamayo} D.,  2018, \mn@doi [\mnras] {10.1093/mnras/stx2479}, \href
  {https://ui.adsabs.harvard.edu/abs/2018MNRAS.473.3351R} {473, 3351}

\bibitem[\protect\citeauthoryear{{Rein}, {Tamayo}  \& {Brown}}{{Rein}
  et~al.}{2019a}]{Reinetal2019b}
{Rein} H.,  {Tamayo} D.,   {Brown} G.,  2019a, \mn@doi [\mnras]
  {10.1093/mnras/stz2503}, \href
  {https://ui.adsabs.harvard.edu/abs/2019MNRAS.489.4632R} {489, 4632}

\bibitem[\protect\citeauthoryear{{Rein}, {Brown}  \& {Tamayo}}{{Rein}
  et~al.}{2019b}]{Reinetal2019c}
{Rein} H.,  {Brown} G.,   {Tamayo} D.,  2019b, \mn@doi [\mnras]
  {10.1093/mnras/stz2942}, \href
  {https://ui.adsabs.harvard.edu/abs/2019MNRAS.490.5122R} {490, 5122}

\bibitem[\protect\citeauthoryear{{Smith}}{{Smith}}{1977}]{Smith77}
{Smith} Jr. H.,  1977, \aap, \href
  {https://ui.adsabs.harvard.edu/abs/1977A%26A....61..305S} {61, 305}

\bibitem[\protect\citeauthoryear{{Tamayo}, {Rein}, {Shi}  \&
  {Hernandez}}{{Tamayo} et~al.}{2019}]{Tamayoetal2019}
{Tamayo} D.,  {Rein} H.,  {Shi} P.,   {Hernandez} D.~M.,  2019, arXiv e-prints,
  \href {https://ui.adsabs.harvard.edu/abs/2019arXiv190805634T} {}

\bibitem[\protect\citeauthoryear{{Urminsky}}{{Urminsky}}{2010}]{Urminsky2010}
{Urminsky} D.~J.,  2010, \mn@doi [\mnras] {10.1111/j.1365-2966.2010.16974.x},
  \href {https://ui.adsabs.harvard.edu/abs/2010MNRAS.407..804U} {407, 804}

\bibitem[\protect\citeauthoryear{{Valtonen}}{{Valtonen}}{1976}]{Valtonen76}
{Valtonen} M.~J.,  1976, \mn@doi [\apss] {10.1007/BF01225963}, \href
  {https://ui.adsabs.harvard.edu/abs/1976Ap%26SS..42..331V} {42, 331}

\bibitem[\protect\citeauthoryear{{Wisdom}}{{Wisdom}}{2006}]{W06}
{Wisdom} J.,  2006, \mn@doi [AJ] {10.1086/500829}, \href
  {http://adsabs.harvard.edu/abs/2006AJ....131.2294W} {131, 2294}

\bibitem[\protect\citeauthoryear{{Wisdom}}{{Wisdom}}{2015}]{W15}
{Wisdom} J.,  2015, \mn@doi [AJ] {10.1088/0004-6256/150/4/127}, \href
  {http://adsabs.harvard.edu/abs/2015AJ....150..127W} {150, 127}

\bibitem[\protect\citeauthoryear{{Wisdom}}{{Wisdom}}{2018}]{Wisdom2018}
{Wisdom} J.,  2018, \mn@doi [\mnras] {10.1093/mnras/stx2906}, \href
  {https://ui.adsabs.harvard.edu/abs/2018MNRAS.474.3273W} {474, 3273}

\bibitem[\protect\citeauthoryear{{Wisdom} \& {Hernandez}}{{Wisdom} \&
  {Hernandez}}{2015}]{WH15}
{Wisdom} J.,  {Hernandez} D.~M.,  2015, \mn@doi [MNRAS]
  {10.1093/mnras/stv1862}, \href
  {http://adsabs.harvard.edu/abs/2015MNRAS.453.3015W} {453, 3015}

\bibitem[\protect\citeauthoryear{{Wisdom} \& {Holman}}{{Wisdom} \&
  {Holman}}{1991}]{WH91}
{Wisdom} J.,  {Holman} M.,  1991, \mn@doi [AJ] {10.1086/115978}, 102, 1528

\bibitem[\protect\citeauthoryear{{Wisdom}, {Holman}  \& {Touma}}{{Wisdom}
  et~al.}{1996}]{WHT96}
{Wisdom} J.,  {Holman} M.,   {Touma} J.,  1996, Fields Institute
  Communications, Vol.~10, p.~217, \href
  {http://adsabs.harvard.edu/abs/1996FIC....10..217W} {10, 217}

\bibitem[\protect\citeauthoryear{{Yoshida}}{{Yoshida}}{1990}]{Y90}
{Yoshida} H.,  1990, Physics Letters A, 150, 262

\makeatother
\end{thebibliography}
\appendix
\section{{Initial conditions}}
\label{sec:init}
{
The initial conditions used for the test problem of Section \ref{sec:methods} are,
\begin{equation}
\label{eq:icscale}
\begin{aligned}
\v{r}_1 &= (-8.810339758984251\times 10^{-6}, 0),\\
\v{r}_2 &= (0.2936779919661417,0), \\ 
\v{r}_3 &=  (0.2505482770417838, 0), \\
\v{v}_1 &= (0, -5.535699732490998\times 10^{-5}), \\
\v{v}_2 &= (0, 1.8452332441636659), \qquad \mathrm{and} \\
\v{v}_3 &= (4.921004910368913\times 10^{-16}, 1.7962485792165686). 
\end{aligned}
\end{equation}
The positions and velocities have format ($x$-coordinate, $y$-coordinate).  The exact initial conditions used from the test problem of Section \ref{sec:choosing} are given {by},
\begin{equation}
\label{eq:ic}
\begin{aligned}
\v{r}_1 &= ( -8.810339758984251\times 10^{-6}, 0),\\
\v{r}_2 &= (0.2936779919661417, 0), \\ 
\v{r}_3 &= (0.28307962227403155, -6.93345106643469\times 10^{-17}), \\
\v{v}_1 &= ( 0, -5.535699732490998\times 10^{-5}), \\
\v{v}_2 &= (0, 1.8452332441636659), \qquad \mathrm{and} \\
\v{v}_3 &= (4.604340604363928\times 10^{-16}, 1.8450746882058686). 
\end{aligned}
\end{equation}
Note most initial conditions are the same as eqs. \eqref{eq:icscale}.
}
\section{Past $N$-body reliability studies}
\label{sec:lit}
\cite{Valtonen76,Smith77,Heggie91,PB14,BP15} have used conventional integrators\footnote{Symplectic integrators could also be considered, see \cite{HB15,H16,DH17}} to solve $N$-body systems and then calculate satistics on the various systems.  \cite{Valtonen76,PB14,BP15} study three-body problems.  \cite{Valtonen76} runs scattering experiments of a particle scattering of a  binary, with a variety of initial conditions, finding the distribution of outcomes does not depend on the integration accuracy.  \cite{PB14,BP15} run democratically interacting three-body problems and measure statistics such as the dissolution time.  They compare against an arbitrary precision code to find the right statistics are produced if energy conservation is enforced to at least $1$ part in $10$.  Their integrations included both short and long-term results.  \cite{Smith77} studies $16$-body problems for up to roughly $17$ e-folding times--- we describe calculation of this number in Appendix \ref{sec:efold}.  \cite{Smith77} found convergence of statistics such as the half-mass radius as long as the energy error was conserved better than about $10^{-1}$ after the $17$ e-folding times.  \cite{Heggie91} studies $N = 100$ problems until the first hard binary formed after about $23$ e-folding times (Appendix \ref{sec:efold}).  He showed that the half-mass radius converged as the accuracy increased, but, alarmingly, not all statistics converge.  The statistic of the number of escaped particles and their energies did not converge.  We warn that in these studies, convergence does not guarantee an accurate statistic, inserting another difficulty into $N$-body studies.

\section{{\Large \lowercase{e}}-folding times}
\label{sec:efold}
Here, we calculate the number of e-folding times in the self-gravitating $N$-body simulations of \cite{Heggie91} and \cite{Smith77}.  We use the relationship between the relaxation and crossing times \citep[Eq. (1.38)]{BT08}:
\begin{equation}
t_{\mathrm{relax}} \approx \frac{0.1 N}{\ln N} t_{\mathrm{cross}}.
\label{eq:relax}
\end{equation}
\citeauthor{Heggie91}'s $N = 100$ runs were stopped after the first energetic binary formed, at approximately $t = 32$, in $N$-body units \citep{Henon1971}.  Using the crossing time of $2 \sqrt{2}$, and the fact than an e-folding time is roughly the crossing time \citep{GHH93}, it appears $23$ e-folding times were run.  But this estimate does not take into account that the core crossing time decreases during core collapse, actually substantially increasing the number of e-folding times.  

\citeauthor{Smith77}'s $N = 16$ runs were integrated for a maximum of $10$ relaxation times.  Using \eqref{eq:relax}, this translates to roughly $17$ e-folding times.  However, this number may not be accurate since \citeauthor{Smith77} ran some integrations with low accuracy and a small number of stored significant digits, possibly violating the assumptions of \citep{GHH93}.

\end{document}